\documentclass[10pt,conference]{IEEEtran}
\IEEEoverridecommandlockouts  

\usepackage{main_macro}

\begin{document}

\title{Understanding Data Science Lifecycle Provenance via Graph Segmentation and Summarization
}

\author{
  \IEEEauthorblockN{Hui Miao\IEEEauthorrefmark{1},~ Amol Deshpande}
  \thanks{\IEEEauthorrefmark{1}Work done at UMD, now at Google.}
  \IEEEauthorblockA{\textit{Department of Computer Science}, \\
  \textit{University of Maryland}, College Park, MD 20742\\
  {\large{\{}}hui,~amol{\large{\}}}@cs.umd.edu}
}

\maketitle

\begin{abstract}
Along with the prosperous data science activities, the importance of provenance during data science project lifecycle is recognized and discussed in recent data science systems research. 
Increasingly modern data science platforms today have non-intrusive and extensible provenance ingestion mechanisms to collect rich provenance and context information, handle 
modifications to the same file using distinguishable versions, and use {\em graph} data models (e.g., property graphs) and query languages (e.g., Cypher) to represent and manipulate
the stored provenance/context information. \eat{However, d}Due to the schema-later nature of the metadata, multiple versions of the same files, and unfamiliar artifacts introduced by team members\eat{, and
enormous provenance records collected continuously}, the ``provenance graph'' is verbose and evolving, and hard to understand\eat{ for the users}; \eat{just }using standard graph
query model, it is \eat{very }difficult to compose queries and utilize this valuable information. 

In this paper, we propose two high-level graph query operators to address the verboseness and evolving nature of such provenance graphs. 
First, we introduce a {\bf graph segmentation} operator, which queries the \emph{retrospective provenance} between a set of source vertices\eat{ (e.g., today's \emph{dataset})} and a set of destination vertices\eat{ (e.g.,
        today's \emph{result})} via flexible boundary criteria to help users get insight about the derivation relationships among those vertices. 
We show the semantics of such a query in terms of a context-free grammar, and develop efficient\eat{ context-free reachability} algorithms that run orders of magnitude faster than state-of-the-art. 
Second, we propose a {\bf graph summarization} operator that combines similar segments together to query \emph{prospective provenance} of the underlying project\eat{ (e.g., daily pipeline between \emph{dataset} and \emph{result})}. The operator allows 
tuning the summary 
by ignoring vertex details and characterizing local structures, and ensures the provenance meaning 
using path constraints. We show the optimal summary problem is PSPACE-complete and develop effective approximation algorithms.
The operators are implemented\eat{ in our \provdb\ system} on top of a property graph backend\eat{ (Neo4j)}. 
We evaluate our query methods extensively\eat{ on a variety of synthetic provenance graphs by mimicking real-world project behaviors} and show the effectiveness and efficiency of the proposed methods.

\end{abstract}

\section{Introduction}
Provenance capture and analysis is being increasingly seen as a crucial enabler for prosperous data science activities~\cite{ground@cidr17,provdb@hilda17,google_tutorial@sigmod17,amazon_metadata@learningsys17,modeldb_HILDA16p}. 
\eat{In general, c}Capturing provenance allows the practitioners introspect the data analytics trajectories, monitor the ongoing modeling activities, increase auditability, aid in reproducibility, and communicate the practice with others~\cite{provdb@hilda17}. Specific systems have been developed to help diagnose \eat{distributed}dataflow programs~\cite{lipstick@pvldb11,titian@pvldb15}, 
ingest provenance \eat{during}in the lifecycle~\cite{modeldb_HILDA16p,provdb@hilda17}, and \eat{understand}manage pipelines for high-level modeling paradigms~\cite{modelhub@icde17,franklin@hpdc17}. 

Compared with well-established data provenance systems for databases~\cite{survey_chiew@ftdb09}, and scientific workflow systems for e-science~\cite{workflow_survey@cse08}, building provenance systems for data science faces an \emph{unstable} data science lifecycle that is often ad hoc, typically featuring highly unstructured datasets, an amalgamation of different tools and techniques, significant back-and-forth among team members, and trial-and-error to identify the right analysis tools, models, and parameters. 
Schema-later approaches and graph data model are often used to capture the lifecycle, versioned artifacts and associated rich information~\cite{ground@cidr17,provdb@hilda17,amazon_metadata@learningsys17}, which also echoes the modern provenance data model standardization \eat{over a long period of time }as a result of \eat{long period community }consolidation for scientific workflows~\cite{opm_v11@fgcs11} and the Web~\cite{prov_stack@w3c_tr13}. 

Although there is an enormous potential value of data science lifecycle provenance, e.g., reproducing the results or accelerate the modeling process, the evolving and verbose nature of the
captured provenance graphs makes them difficult to store and manipulate. Depending on the granularity, storing the graphs could take dozens of GBs within several minutes~\cite{linuxprov_abates@atc15}.
More importantly, the verbosity and diversity of the provenance graphs makes it difficult to write general queries to explore and utilize them; there are often no predefined workflows, i.e., the pipelines
change as the project evolves, and instead we have arbitrary steps (e.g., trial and error) in the modeling process. In addition, though storing the provenance graph in a graph database seems like a
natural choice, most of the provenance query types of interest involve paths~\cite{prov_challenges@website}, and require returning paths instead of answering yes/no queries like
reachability~\cite{pql_harvard@ipaw08}. Writing queries to utilize the lifecycle provenance is beyond the capabilities of the pattern matching query (BPM) and regular path query (RPQ) support in popular graph databases~\cite{gquery_theory@pods13,survey_querylang@arix16,pgql_oracle@grades16}.
For example, answering `how is today's result file generated from today's data file' requires a segment of the provenance graph that includes not only the mentioned files but also other files that are not
on the path that the user may not know at all (e.g., `a configuration file'); answering `how do the team members typically generate the result file from the data file?' requires summarizing several query results of the above query while keeping the result meaningful from provenance perspective.

Lack of proper query facilities in modern graph databases not only limits the value of lifecycle provenance systems for data science, but also of other provenance systems. 
The specialized query types of interest in the provenance domain~\cite{prov_challenges@website,pql_harvard@ipaw08} had often led provenance systems to implement specialized storage
systems~\cite{pass_harvard@atc06,compression_bertram@edbt09} and query interfaces~\cite{zoom_penn@icde08,bertram_lang@edbt10} on their own~\cite{workflow_survey@cse08}. Recent works \eat{on provenance
    graphs }in the provenance community propose various graph transformations for different tasks, which are essentially different template queries from the graph querying perspective; these include
    grouping vertices together to handle publishing policies~\cite{provabs_pmissier@ipaw14}, aggregating vertices in a verbose graph to understand commonalities and outliers~\cite{agg_lucmoreau@gam15},
    segmenting a provenance graph \eat{via declarative language }for feature extractions in cybersecurity~\cite{prov_segmentation@tapp16}. {\em Our goal with this work is to initiate a more systematic study of
    abstract graph operators that modern graph databases need to support to be a viable option for storing provenance graphs.} 

Toward that end, in this paper, we propose two graph operators for common provenance queries to let the user explore the evolving provenance graph without fully understanding the underlying provenance graph structure. The operators not only help our purpose in the context of data science but also other provenance applications which have no clear workflow skeletons and are verbose in nature (e.g., ~\cite{linuxprov_abates@atc15,agg_lucmoreau@gam15,prov_segmentation@tapp16}).

First, we introduce a flexible graph segmentation operator for the scenario when users are not familiar with the evolving provenance graph structure but still need to query 
\eat{It queries the }\emph{retrospective provenance} between a set of source vertices (e.g., today's data file) and a set of destination vertices (e.g.,
        today's result file) via certain boundary criteria (e.g., hops, timestamps, authorship). The operator is able to induce vertices that contribute to the destination vertices in a similar way w.r.t.
the given source vertices in the specified boundary. Parts of the segmentation query require a context-free language (CFL) to express its semantics. We study how to support such CFL query in
provenance graphs, exploit novel grammar rewriting schemes and propose evaluation techniques that run orders of magnitude faster than state-of-the-art for our graph operator. 

Second, we propose a graph summarization operator for aggregating the results of segmentation operations, in order to analyze the \emph{prospective provenance} of the underlying project (e.g., typical pipeline from the data file to the result file). It allows the user to tune the summary graph 
by ignoring vertex details and characterizing local structures during aggregation, and ensures the summary is meaningful from the provenance perspective through path constraints. We show the optimal
summary problem is PSPACE-complete and develop effective approximation algorithms that obey the path constraints.

We illustrate the operators on provenance data model standards (W3C \prov); 
the formulations and evaluation techniques are general to many provenance-aware applications. We show how to build such operators on top of modern property graph backends (Neo4j). We present extensive experiments that show the effectiveness and efficiency of our proposed techniques on synthetic provenance graph datasets mimicking real-world data science projects.

\eat{
Our key contributions are as follows:
\begin{list}{$\bullet$}{\leftmargin 0.15in \topsep 0pt}
  \item We introduce the segmentation operation for the scenario when users are not familiar with the evolving provenance graph structure but still need to query retrospective provenance. We are the first
  to use a context-free language as a provenance query primitive and develop efficient evaluation algorithms for provenance graphs.
  \item We propose the summarization operation that combines similar  segments and preserves provenance meaning for querying prospective provenance in lifecycles without predefined workflow skeletons.
  We analyze its complexity and develop efficient algorithms.
  \item We present extensive experiments that show the effectiveness and efficiency of our proposed techniques on synthetic provenance graph datasets mimicking real-world data science projects.
  \item The provenance queries, their formulations and evaluation techniques are general to many provenance-aware applications. We show how to build a system on top of modern graph databases.
\end{list}

In the rest of the paper, we first describe provenance management for data science lifecyles, by introducing the provenance model and unique challenges \eat{of the query type of interests }in Sec.~\ref{sec:background}. Then we describe the segmentation operation and summarization operation \eat{respectively }in Sec.~\ref{subsec:seg_op} and Sec.~\ref{subsec:sum_op}. The implementation of the operations are discussed in Sec.~\ref{sec:system}, followed by the extensive experiments in Sec.~\ref{sec:exp}. We summarize the related work from different communities in Sec.~\ref{sec:related_work}, and conclude with future directions in Sec.~\ref{sec:conclusion}.
}

\topic{Outline}: We begin with an overview of our system, introduce the provenance model and unique challenges (Sec.~\ref{sec:background}), then describe semantics and evaluation algorithms for segmentation (Sec.~\ref{subsec:seg_op}) and summarization operators (Sec.~\ref{subsec:sum_op}). We present extensive experimental evaluation (Sec.~\ref{sec:exp}) and summarize the related work from different communities (Sec.~\ref{sec:related_work}).

\begin{figure}[t!]
\centering{
\includegraphics[width=0.42\textwidth]{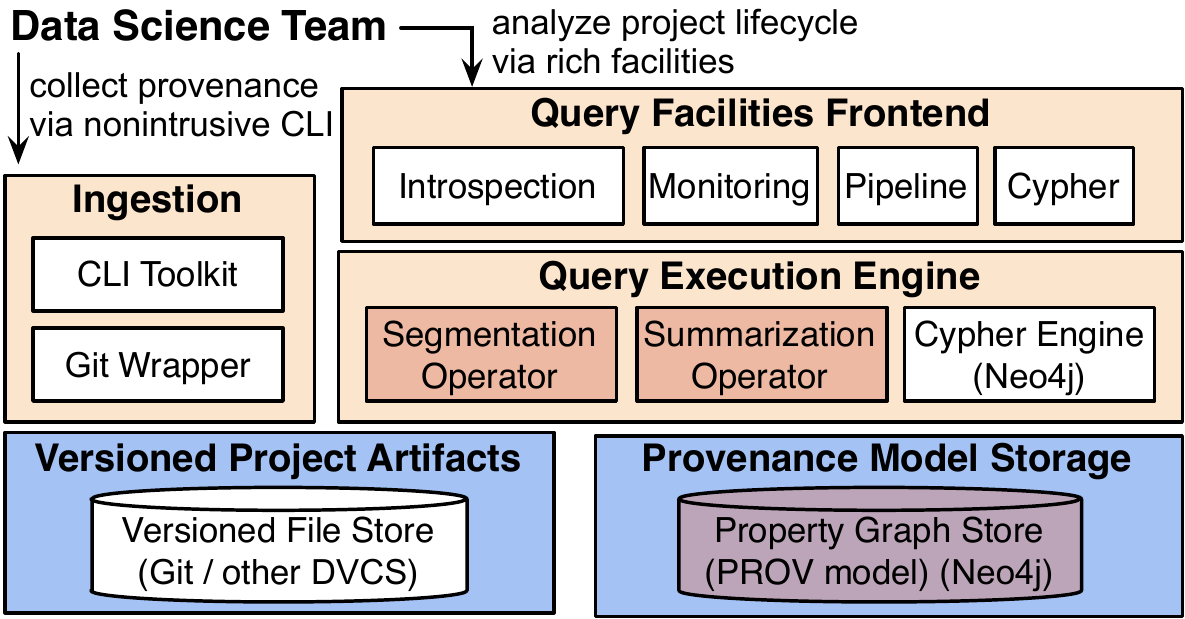}
}
\vspace{-5pt}
\caption{Overview of a Lifecycle Management System, \provdb~\cite{provdb@hilda17}}
\label{fig:sys_arch}
\vspace{-12pt}
\end{figure}

\begin{figure*}[t!]
\centering{
\subfigure[An Example for a Data Science Project Lifecycle \& Associated Provenance]{
  \includegraphics[width=0.645\textwidth]{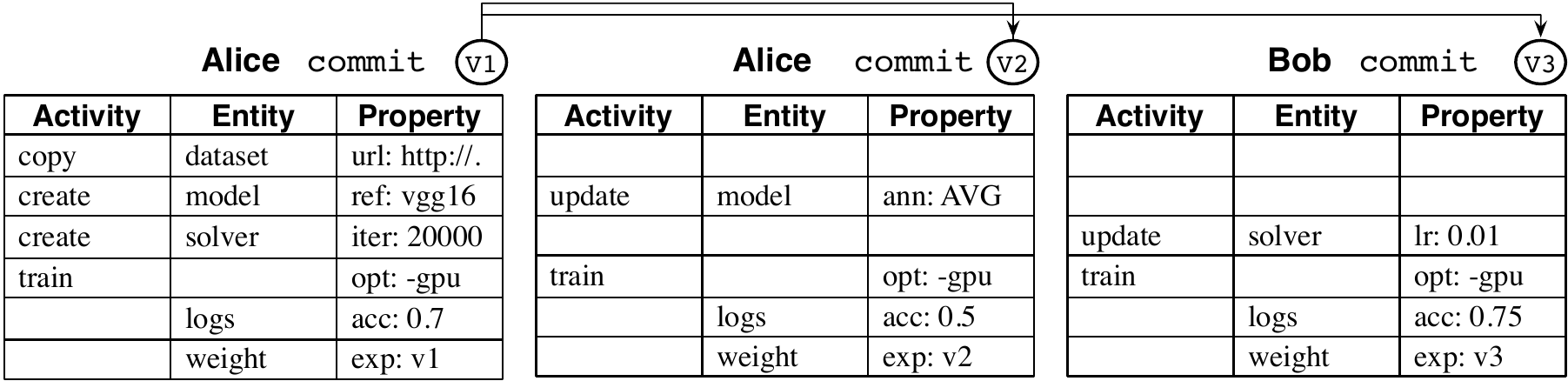}
  \label{fig:example_lifecycle}
}
\subfigure[Illustration of the W3C \prov\ Data Model]{
  \includegraphics[width=0.32\textwidth,trim=0 0 0 10]{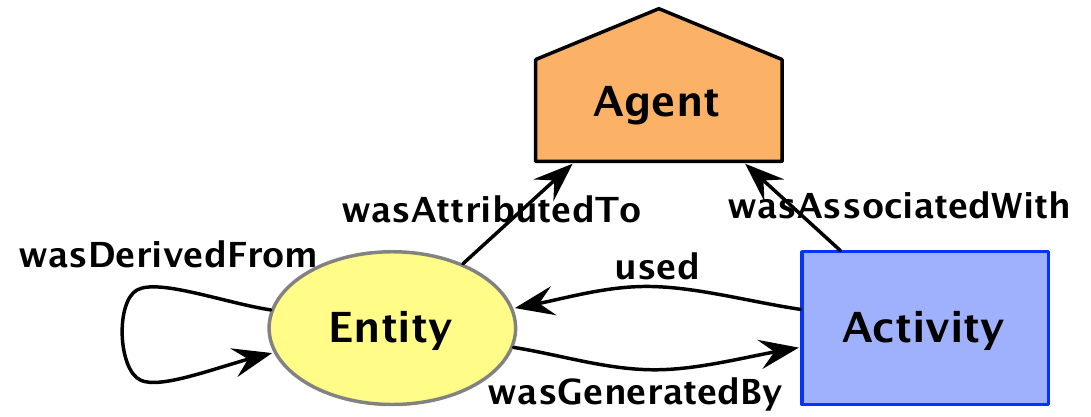}
  \label{fig:prov}
}
\\\vspace{-3.2pt}
\subfigure[Provenance Graph for the Lifecycle Example (Some edges and properties are not shown due to space constraints)]{
  \includegraphics[width=1.0\textwidth,trim=0 0 0 12]{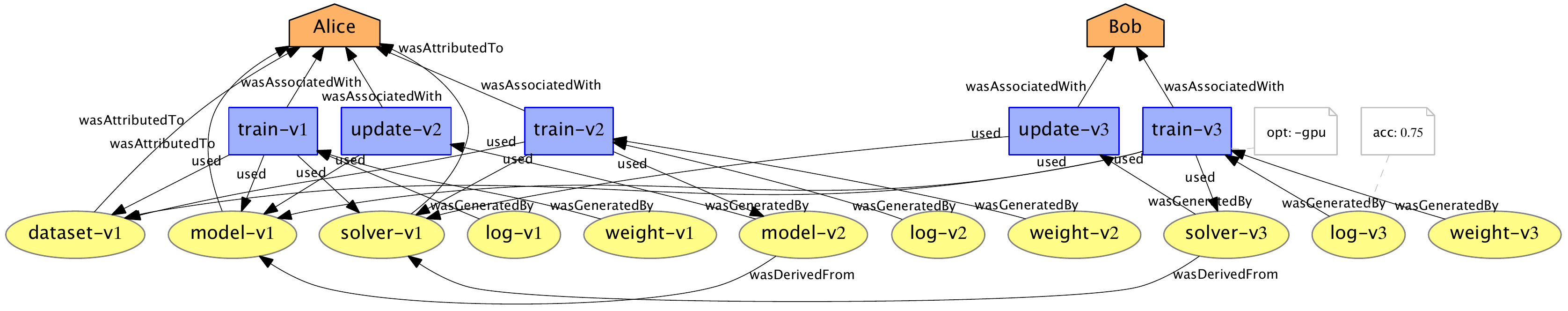}
  \label{fig:example_provenance}
}
\\\vspace{-3.2pt}
\subfigure[Segmentation Query Examples]{
  \includegraphics[width=0.475\textwidth,trim=0 0 0 12]{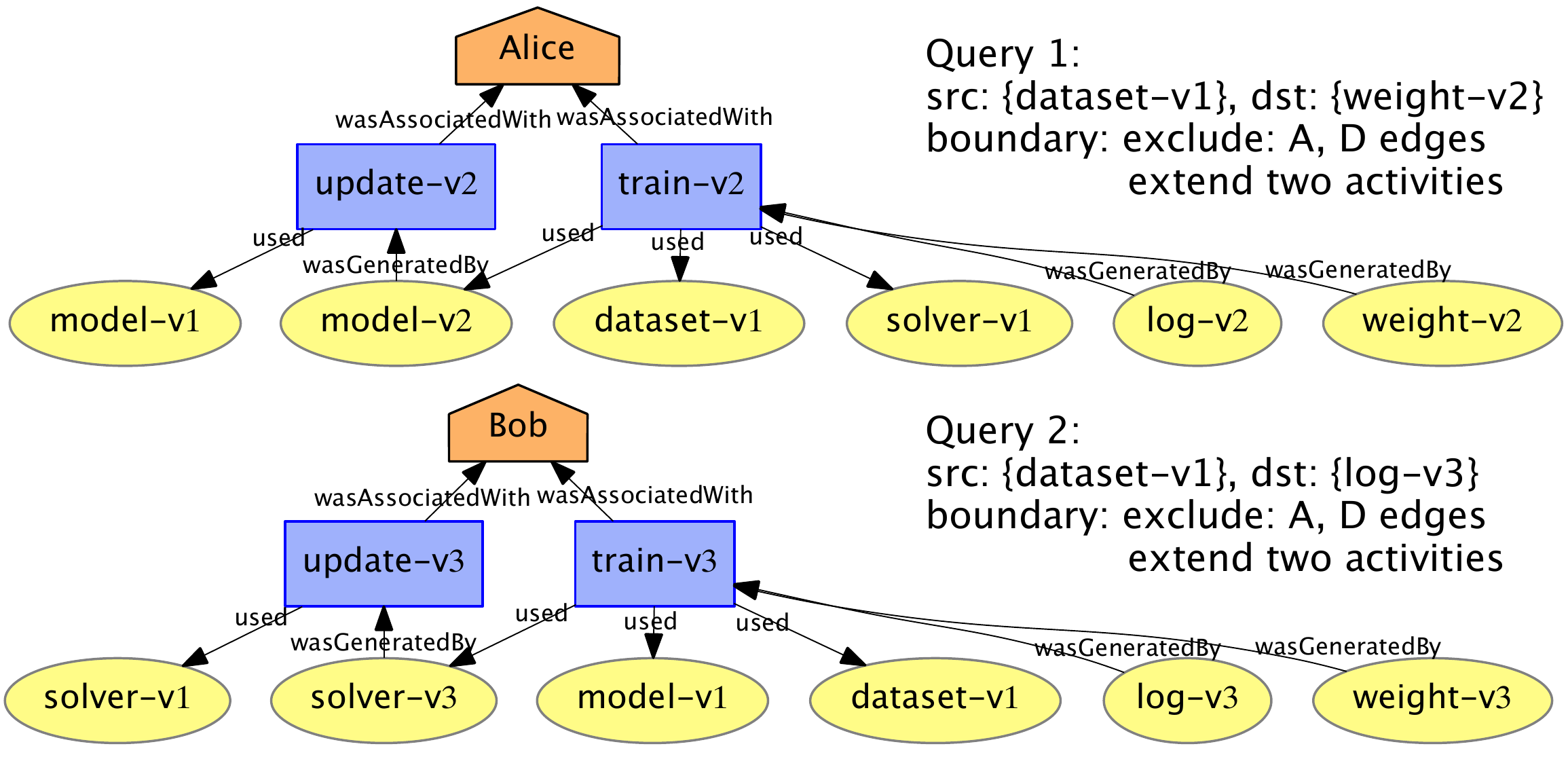}
  \label{fig:example_query_segmentation}
}
\subfigure[Summarization Query Examples]{
  \includegraphics[width=0.48\textwidth,trim=0 0 0 12]{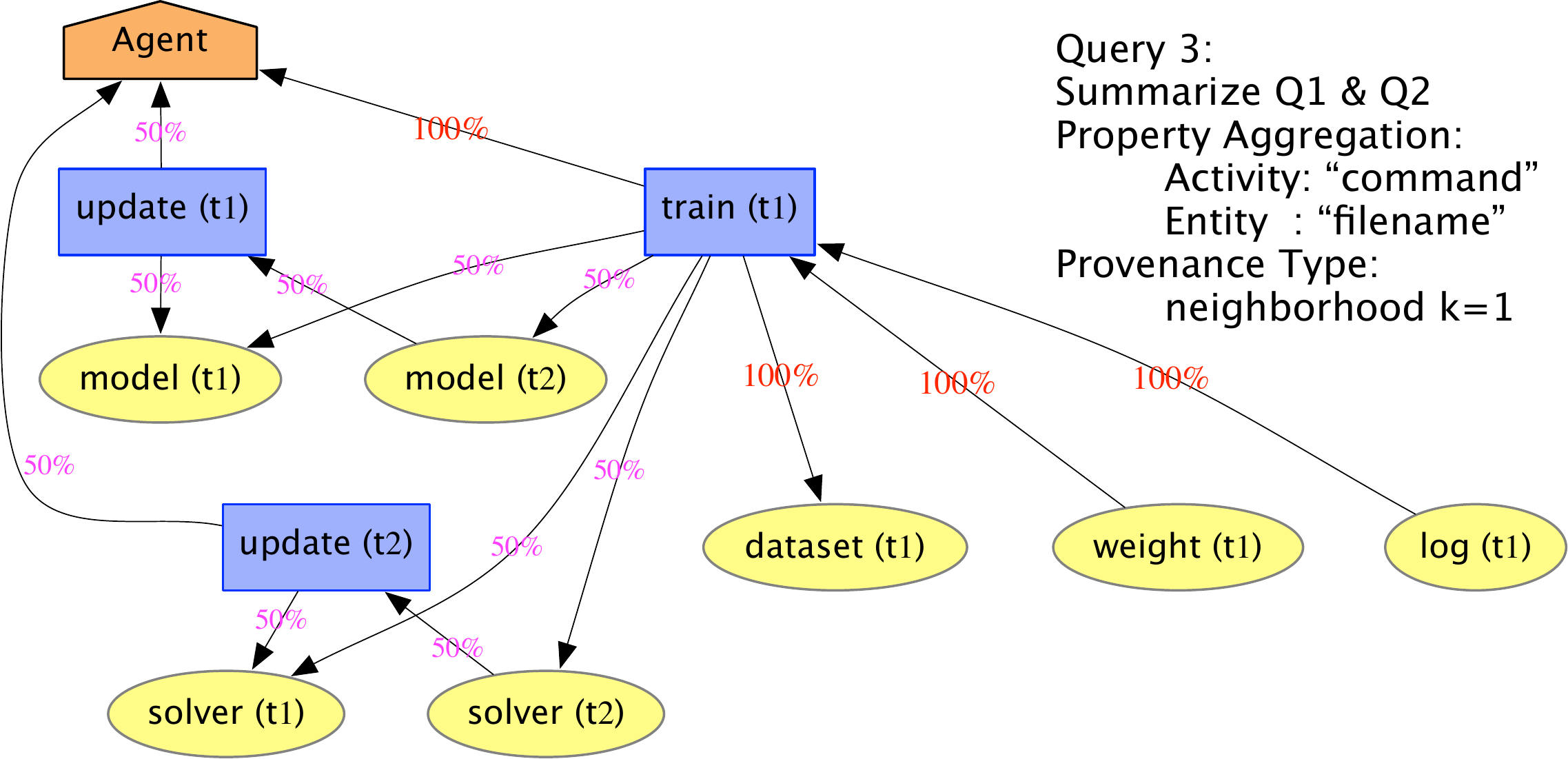}
  \label{fig:example_query_summarization}
}
}
\vspace{-5pt}
\caption{Illustration of Provenance Data Model and Query Operators in Data Science Lifecycles}
\vspace{-12pt}
\label{fig:example}
\end{figure*}

\section{Overview}
\label{sec:background}

We mainly focus on querying provenance for data science lifecycles, i.e., collaborative pipelines consisting of versioned artifacts and derivation and transformation steps, often
parameterized~\cite{ground@cidr17,provdb@hilda17,amazon_metadata@learningsys17}. We show our high-level system architecture and introduce the background with a motivating example. We then define the standard provenance model, and summarize typical provenance query types of interest and analyze them from the perspective of graph querying.

\subsection{System Design \& Motivating Example}

Data science projects can range from well-defined prediction tasks (e.g., predict labels given images) to building and monitoring a large collection of modeling or analysis pipelines, often over a
    long period of time~\cite{google_tutorial@sigmod17,amazon_notebook@vldb16,keystoneml_amp@icde17,modelhub@icde17}. 
Using a lifecycle provenance management system (Fig.~\ref{fig:sys_arch})\eat{~\cite{ground@cidr17,provdb@hilda17}}, details of the project progress, versions of the artifacts and associated provenance are captured and managed. 
In Example~\ref{exp:motivation}, we use a classification task using neural networks to illustrate the system background, provenance model and query type.

\begin{example}
\label{exp:motivation}
In Fig.~\ref{fig:example_lifecycle}, Alice and Bob work together on a classification task to predict face ids given an image. Alice starts the project and creates a neural network by modifying a popular model. She downloads the dataset and edits the model definitions and solver hyperparameters, then invokes the training program with specific command options. After training the first model, she examines the accuracy in the log file, annotates the weight files, then commits a version using \cmd{git}. As the accuracy of the first model is not ideal, she changes the \eat{neural }network  by editing the model definition, trains it again and derives new log files and weight parameters. However the accuracy drops, and she turns to Bob for help. Bob examines what she did, trains a new model following some best practices by editing the solver configuration in version $v_1$, and commits a better model. 

Behind the scene, a lifecycle management system (Fig.~\ref{fig:sys_arch}) tracks user activities, manages project artifacts (e.g., datasets, models, solvers) and ingests provenance. In the Fig.~\ref{fig:example_lifecycle} tables, we show ingested information in detail: \emph{a)} history of user activities (e.g., the first \emph{train} command uses model $v_1$ and solver $v_1$ and generates logs $v_1$ and weights $v_1$), \emph{b)} versions and changes of entities (e.g., weights $v_1$, $v_2$ and $v_3$) and derivations among those entities (e.g., model $v_2$ is derived from model $v_1$), and \emph{c)} provenance records as associated properties to activities and entities, ingested via provenance system ingestors (e.g., dataset is copied from some url, Alice changes a pool layer type to AVG in $v_2$, accuracy in logs $v_3$ is $0.75$).
\end{example} 

\topic{Provenance Model}: 
The ingested provenance of the project lifecycle naturally forms a provenance graph, which is a directed acyclic graph\eat{\footnote{In our system, we use versioning to avoid cyclic self-derivations of the same entity and overwritten entity generations by some activity.}} and encodes information with multiple aspects, such as a version graph representing the artifact changes, a workflow graph reflecting the derivations of those artifact versions, and a conceptual model graph showing the involvement of problem solving methods in the project\eat~\cite{ground@cidr17,provdb@hilda17}. To represent the provenance graph and keep our discussion general to other provenance systems, we choose the W3C \prov\ data model~\cite{prov_dm@w3c_tr13}, which is a standard interchange model for different provenance systems.

\eat{The full \prov\ data model is complex in order to satisfy application needs for different domains~\cite{prov_dm@w3c_tr13}. Due to space constraints and for simplicity, we use the core subset of it, which is shown in Fig.~\ref{fig:prov}. 
}
We use the core set of \prov\ data model shown in Fig.~\ref{fig:prov}. 
There are three types of vertices (\vertexset) in the provenance graph: 
\begin{enumerate}[leftmargin=18pt,label=\emph{\alph*})]{}
\item 
\ul{Entities} (\entity) are the project artifacts (e.g., files, datasets, scripts) which the users work on and talk about{ in a project, and the underlying lifecycle management
    system manages their provenance}; 
\item
\ul{Activities} (\activity) are the system or user actions (e.g., \cmd{train}, \cmd{git commit}, \cmd{cron} jobs) which act upon or with entities over a period of time, $[t_i, t_j)$; 
\item
\ul{Agents} (\agent) are the parties who are responsible for some activity (e.g., a team member, a system component). 
\end{enumerate}
Among vertices, there are five types of directed edges{\footnote{There are 13 types of relationships among Entity, Activity and Agent. The proposed
    techniques in the paper can be extended naturally to support more relation types in other provenance systems.}} (\edgeset): 
\begin{enumerate}[leftmargin=18pt,label={\roman*})]{}
\item    
An activity started at time $t_i$ often uses some entities (\ul{`used'}, \used $\subseteq$ \activity $\times$ \entity);  
\item
then some entities would be generated by the same activity at time $t_j$ ($t_j \geq t_i$) (\ul{`wasGeneratedBy'}, \wasGeneratedBy $\subseteq$ \entity $\times$ \activity); 
\item
An activity is associated with some agent during its period of execution (\ul{`wasAssociatedWith'}, \wasAssociatedWith $\subseteq$ \activity $\times$ \agent);
For instance, in Fig.~\ref{fig:example_lifecycle}, the activity \cmd{train} was associated with Alice, used a set of artifacts (model, solver, and dataset) and generated other artifacts (logs, weights).
In addition: 
\item
Some entity's presence can be attributed to some agent (\ul{`wasAttributedTo'}, \wasAttributedTo $\subseteq$
    \entity $\times$ \agent). e.g., the dataset  was added from external sources and attributed to Alice; 
\item
An entity was derived from another entity (\ul{'wasDerivedFrom'}, \wasDerivedFrom $\subseteq$ \entity $\times$ \entity), such as versions of the same artifact (e.g.,
    different model versions in $v_1$ and $v_2$ in Fig.~\ref{fig:example_lifecycle}). 
\end{enumerate}    

In the provenance graph, both vertices and edges have a label to encode their vertex type in $\{$\entity, \activity, \agent$\}$ or edge type in $\{$\used, \wasGeneratedBy, \wasAssociatedWith, \wasAttributedTo, \wasDerivedFrom$\}$. \eat{All o}Other \ul{provenance records} are modeled as properties, ingested by \eat{a set of configured project ingestors}the system during \eat{the period of }activity executions and represented as key-value pairs.

\eat{\prov\ standard defines various serializations of the concept model\eat{, such as RDF, XML, and JSON} (e.g., RDF, XML, JSON)~\cite{prov_stack@w3c_tr13}. In our system, we use a physical property graph data model to store it, as it is more natural for the users to think the artifacts as nodes when writing queries using Cypher or Gremlin. It is also more compact than RDF graph for the large amount of provenance records, which are treated as literal nodes. We discuss implementation details in Sec.~\ref{sec:system}. 
As a summary, we formally define the provenance graph used in the rest of the paper.}

\begin{definition}[Provenance Graph]
\label{def:provgraph}
Provenance in a data science project is represented as a directed acyclic graph (DAG), $\text{\provgraph}(\text{\vertexset}, \text{\edgeset}, \text{\vlabelfunc}, \text{\elabelfunc}, \text{\vpropfunc},\text{\epropfunc})$, where vertices have three types, \vertexset $=$ \entity $\cup$ \activity $\cup$ \agent, and edges have five types, \edgeset $=$ \used $\cup$ \wasGeneratedBy $\cup$ \wasAssociatedWith $\cup$ \wasAttributedTo $\cup$ \wasDerivedFrom. Label functions, \vlabelfunc $:$ \vertexset $\mapsto \{$\entity$,\ $\activity$,\ $\agent$\}$, and \elabelfunc $:$ \edgeset $\mapsto \{$\used$,\ $\wasGeneratedBy$,\ $\wasAssociatedWith$,\ $\wasAttributedTo$,\ $\wasDerivedFrom$\}$ are total functions associating each vertex and each edge to its type. In a project, we refer to the set of property types as \property\ and their values as $\mathcal O$, then vertex and edge properties, \vpropfunc $:$ \vertexset $\times $ \property $\mapsto \mathcal O$ and \eat{edge properties }\epropfunc $:$ \edgeset $\times $ \property $\mapsto \mathcal O$, are partial functions from vertex/edge and property type to some value.
\end{definition}

\begin{example}
\label{exp:motivation_in_prov}
Using the \prov\ data model, in Fig.~\ref{fig:example_provenance}, we show the corresponding provenance graph of the project lifecycle listed in Fig.~\ref{fig:example_lifecycle}. 
Vertex shapes follow their type in Fig.~\ref{fig:prov}. 
Names of the vertices (e.g., `model-v1', `train-v3', `Alice') are made by using their representative properties (i.e., project artifact names for entities, operation names for activities, and first names for agents) and suffixed using the version ids to distinguish different snapshots. 
Activity vertices are ordered from left to right w.r.t. the temporal order of their executions.
We label the edges using their types and show a subset of the edges in Fig.~\ref{fig:example_lifecycle} to illustrate usages of five relationship types. 
\eat{
Note there are many snapshots of the same artifact in different versions, and between the versions, we maintain derivation edges `wasDerivedFrom' (\wasDerivedFrom) for efficient versioning storage.} 
The figure shows the provenance of those entities in all three versions. 
The property records are shown as white boxes but not treated as vertices in the property graph. 
\end{example} 

\topic{Characteristics}: The provenance graph has\eat{ the} following  characteristics to be considered\eat{, which we need to consider} when designing query facilities:
\begin{list}{}{\leftmargin 0.02in \topsep 0pt} 

\item $\bullet$ \textbf{Versioned Artifact}: Each entity is a point-in-time snapshot of some artifact in the project. For instance, the query `accuracy of this version of the model' discusses a particular \textbf{\emph{snapshot}} of the model artifact, while `what are the common updates for \emph{solver} before \cmd{train}' refer to the \textbf{\emph{artifact}} but not an individual snapshot. 
\ul{R1: The query facilities need to support both aspects in the graph.} 

\item $\bullet$ \textbf{Evolving Workflows}: Data science lifecycle is explorative and collaborative in nature, so \emph{there is no static workflow skeleton, and no clear boundaries for individual runs} in contrast with workflow systems~\cite{workflow_survey@cse08}. For instance, the modeling methods may change (e.g., from SVM to neural networks), the data processing steps may vary (e.g., split, transform or merge data files), and the user-committed versions may be mixed with code changes, error fixes, thus may not serve as boundaries of provenance queries for entities. 
\ul{R2: The query facility for snapshots should not assume workflow skeleton and should allow flexible boundary conditions.}

\item $\bullet$ \textbf{Partial Knowledge in Collaboration}: Each team member may work on and be familiar with a subset of artifacts and activities, and may use different tools or approaches, e.g., in Example~\ref{exp:motivation}, Alice and Bob use different approaches to improve accuracy. 
When querying retrospective provenance of the snapshots attributed to other members or understanding activity process over team behaviors, the user may only have partial knowledge at query time, thus may find it difficult to compose the right graph query. 
\ul{R3: The query facility should support queries with partial information reflecting users' understanding and induce correct result.}

\item $\bullet$ \textbf{Verboseness for Usage}: In practice, the provenance graph would be very verbose for humans to use and in large volume for the system to manage. 
\ul{R4: The query facility should be scalable to large graph and process queries efficiently.}
\end{list}

\subsection{Provenance Query Types of Interest}
\label{subsec:querymodel}
However, the provenance standards (e.g., \prov, {\sc OPM}) do not describe query models, as different systems have their own application-level meaning of those nodes~\cite{prov_stack@w3c_tr13,opm_v11@fgcs11}. 
General queries (e.g., SQL, XPATH, SPARQL) provided by a backend DBMS to express provenance retrieval tend to be\eat{ queries are} very
complex~\cite{workflow_survey@cse08}. To improve usability, a few systems provide novel query
facilities~\cite{visualization@vis05,zoom_penn@icde08}, and some of them propose special query
languages~\cite{pql_harvard@ipaw08,bertram_lang@edbt10}. Recent provenance systems which adopt W3C \prov\ data model naturally use graph
stores as backends~\cite{linuxprov_abates@atc15,websiteprov_abates@www16}; since the standard graph query languages often cannot satisfy the needs~\cite{pql_harvard@ipaw08,gquery_theory@pods13}, a set of graph manipulation techniques is often proposed to utilize the provenance~\cite{provabs_pmissier@ipaw14,agg_lucmoreau@gam15,prov_segmentation@tapp16}.
By observing the characteristics of the provenance graph in analytics lifecycle and identifying the requirements for the query facilities\eat{ (Sec.~\ref{subsec:datamodel})}, \ul{we propose two graph
    operators (i.e., segmentation and summarization)} for general provenance \eat{graphs in \prov\ }data model, that we illustrate next with examples, and discuss in more depth in the next section. 

\topic{Segmentation:} A very important provenance query type of interest is querying ancestors and descendants of entities~\cite{workflow_survey@cse08,pql_harvard@ipaw08}.
In our context, the users introspect the lifecycle and identify issues by analyzing dependencies among snapshots. Lack of a workflow
skeleton and clear boundaries, makes the queries over the provenance graph more difficult. Moreover\eat{ Also} the user may not be
able to specify all interested entities in a query due to partial knowledge. 
We propose a segmentation operator that\eat{ allows the user to specify} takes sets of source and destination entities, and the operator induces other important unknown entities 
satisfying a set of specified boundary criteria.

\begin{example}
\label{exp:segop_q1_2}
In Fig.~\ref{fig:example_query_segmentation}, we show two examples of provenance graph segmentation query. In Query 1 ($Q_1$), Bob was interested in what Alice did in version $v_2$. He did not know the details of activities and the entities Alice touched, instead he set \{\emph{dataset}\}, \{\emph{weight}\} as querying entities to see how the \emph{weight} in Alice's version $v_2$ was connected to the \emph{dataset}. He filtered out uninterested edge types (.e.g, \wasAttributedTo, \wasDerivedFrom) and excluded actions in earlier commits (e.g., $v_1$) by setting the boundaries as two activities away from those querying entities. \eat{In the figure, t}The system found connections among the querying entities, and included vertices within the boundaries. After interpreting the result, Bob knew Alice updated the model definitions in \emph{model}. On the other hand, Alice would ask query to understand how Bob improved the accuracy and learn from him. In Query 2 ($Q_2$), instead of learned \emph{weight}, accuracy property associated \emph{log} entity is used as querying entity along with \emph{dataset}. The result showed Bob only updated solver configuration and did not use her new \emph{model} committed in $v_2$. 
\end{example}

\topic{Summarization:} In workflow systems, querying the workflow skeleton (aka prospective provenance) is an important use case (e.g.,
        business process modeling~\cite{bpql_milo@vldb06}) and included in the provenance challenge~\cite{prov_challenges@website}. In our context, even though a static workflow skeleton is not present, summarizing the skeleton of similar processing pipelines, showing commonalities and identifying abnormal behaviors are very useful query capabilities. However, general graph summarization techniques~\cite{summary_tutorial@pvldb17} are not applicable to provenance graphs due to the subtle provenance meanings and constraints of the data model~\cite{prov_constraints@w3c_tr13,provabs_pmissier@ipaw14,agg_lucmoreau@gam15}. 
We propose a summarization operator \eat{with multi-resolution capabilities for provenance graphs. T}to support querying the artifact
aspect \eat{different aspects }of the provenance. 
\eat{It operates over query results of segmentation 
and allows tuning the summary by ignoring vertex details and characterizing local structures, and ensures provenance meaning through path constraints.
}

\begin{example}
\label{exp:sumop_q3}
\eat{We show a summarization query example in}In Fig.~\ref{fig:example_query_summarization}, an outsider to the team (e.g., some auditor, new team member, or project manager) wanted to understand the activity overview in the project. Segmentation queries (e.g., $Q_1$, $Q_2$ in Fig.~\ref{fig:example_query_segmentation}) only show individual trails of the analytics process at the snapshot level. The outsider issued a summarization query, Query 3 ($Q_3$), by specifying the aggregation over three types of vertices (viewing Alice and Bob as an abstract team member, ignoring details of files and activities), and defining the provenance meanings as a 1-hop neighborhood. The system merged $Q_1$ and $Q_2$ into a summary graph\eat{according to the query}. In the figure, the vertices suffixed name with provenance types to show alternative generation process, while edges are labeled with their frequency of appearance among segments. The query issuer would change the query conditions to derive various summary at different resolutions.
\end{example}

\section{Segmentation Operation}
\label{subsec:seg_op}

Among the snapshots, collected provenance graph describes important ancestry relationships which form `the heart of provenance data'~\cite{pql_harvard@ipaw08}. Often lineages w.r.t. a query or a run graph trace w.r.t. a workflow are used to formulate \eat{such}ancestry queries in relational databases or scientific workflows~\cite{bertram_lang@edbt10}. However, in our context, there are no clear boundaries of logical runs, or query scopes to cleanly define the input and the output. Though a provenance graph could be collected, the key obstacle is lack of formalisms to analyze the verbose information. 
In similar situations for querying script provenance~\cite{ingestions_noworkflow@ipaw14}, Prolog was used to traverse graph imperatively,
   which is an overkill and entails an additional skill-set for team members. \eat{In our system, w}We design \opseg\ to let the users who may only have partial knowledge to query retrospective provenance. \opseg\ semantics induce a connected subgraph to show the ancestry relationships (e.g., lineage) among the entities of interest and include other causal and participating vertices within a \eat{specified }boundary that is adjustable by the users. \eat{Next we first define the elements of the operator and query semantics, followed by query evaluation techniques.}

\subsection{Semantics of Segmentation ({\opseg})}
\label{subsubsec:seg_qelements}

\eat{At a high level, we view t}The \opseg\ operator is a 3-tuple query $(\text{\segSource},\text{\segDestination},\text{\boundaryCriteria})$ on a provenance graph \provgraph\ asking how a set of \ul{\emph{source entities}} \segSource\ $\subseteq$ \entity\ are involved in generating a set of \ul{\emph{destination entities}} \segDestination\ $\subseteq$ \entity. \opseg\ induces \ul{\emph{induced vertices}}\eat{ (entities, activities and agents)} \segInduced\ $\subseteq$ \vertexset\ that show the detailed generation process and satisfy certain \ul{\emph{boundary criteria}} \boundaryCriteria. It returns a connected subgraph \segsubgraphFull\ $\subseteq$ \provgraph, where \segv$=$\segSource$\cup$\segDestination$\cup$\segInduced, and \sege$=$\edgeset\ $\cap$ \segv$\times$\segv.

When discussing the elements of \opseg\ below, we use the following notations for paths in \provgraph. 
A \textbf{\emph{path}} \segpath{$v_0$}{$v_n$}\ connecting vertices $v_0$ and $v_n$ is a vertex-edge alternating sequence $\langle v_0,e_1,v_1,\cdots,v_{n-1},e_{n},v_{n}\rangle$, where $n > 1$, $\forall{i\in[0,n]}\ v_i\in$\ \vertexset, and $\forall{j\in(0,n]}\ e_j=(v_{j-1}, v_j) \in$\ \edgeset. 
Given a path \segpath{$v_0$}{$v_n$}, we define its \textbf{\emph{path segment}} \segsubpath{$v_0$}{$v_n$} by simply ignoring $v_0$ and $v_n$ from the beginning and end of its path sequence, i.e., $\langle e_1,v_1,\cdots,v_{n-1},e_{n}\rangle$.
A \textbf{\emph{path label function}} \pathlabel\ maps a path \segpathshort\ or path segment \segsubpathshort\ to a word by concatenating labels of the elements in its sequence order. Unless specifically mentioned, the label of each element (vertex or edge) is derived via \vlabelfunc$(v)$ and \elabelfunc$(e)$.
For example, from $a$ to $c$, there is a path \segpath{a}{c}$=\langle a,e_a,b,e_b,c\rangle$, where $a, c \in$\entity, $b\in$\activity, $e_a\in$\wasGeneratedBy\ and $e_b\in$\used; its path label \pathlabel$($\segpath{a}{c}$)=$\pathlabelword{\entity\wasGeneratedBy\activity\used\entity}, and its path segment label \pathlabel$($\segsubpath{a}{c}$)=$\pathlabelword{\wasGeneratedBy\activity\used}.
\eat{}
For ease of describing path patterns, for ancestry edges (used, wasGeneratedBy), i.e., $e_k=(v_i,v_j)$ with label \elabelfunc$(e_k)=$\used\ or \elabelfunc$(e_k)=$\wasGeneratedBy, we introduce its virtual \emph{inverse edge} \inv{$e_k$}$=(v_j,v_i)$ with the inverse label \elabelfunc$($\inv{$e_k$}$)=$\usedInv or \elabelfunc$($\inv{$e_k$}$)=$\wasGeneratedByInv\ respectively. A \textbf{\emph{inverse path}} is defined by reversing the sequence, e.g., \inv{\segpath{a}{c}}$=\langle c,$\inv{$e_b$}$,b,$\inv{$e_a$}$,a\rangle$, while \pathlabel$($\inv{\segpath{a}{c}}$)=$\pathlabelword{\entity\usedInv\activity\wasGeneratedByInv\entity}, \pathlabel$($\inv{\segsubpath{a}{c}}$)=$\pathlabelword{\usedInv\activity\wasGeneratedByInv}. 

\eat{Next we discuss \opseg\ semantics and our rationale in detail.}

\subsubsection{Source (\segSource) \& Destination Entities (\segDestination)}
Provenance is about the entities.
In a project, the user know the committed snapshots (e.g., data files, scripts\eat{, and their metadata}) better than the detailed processes generating them. 
When writing a \opseg\ query, we assume the user believes \segSource\ may be ancestry entities of \segDestination. Then \opseg\ reasons their connectivity \eat{among \segSource\ and \segDestination, }and shows other vertices and the generation process which the user may not know and be able to write query with.
Note that users may not know the existence order of entities either, so we allow \segSource\ and \segDestination\ to overlap, and even be identical. In the latter case, the user could be a program~\cite{prov_segmentation@tapp16} and not familiar with the generation process at all. 

\subsubsection{Induced Vertices \segInduced}
Given \segSource\ and \segDestination, 
intuitively \segInduced\ are the vertices
\eat{ (entities, activities and agents)} contributing to the generation process. 
{\bf What vertices should be in \segInduced\ is the core question to ask}. It should reflect the generation process precisely and \eat{better }concisely \eat{in order }to assist the user introspect part of the generation process and make decisions. 

Prior work on inducing subgraphs from a set of vertices do not fit our needs. First, lineage query would generate all ancestors of \segDestination, which is not concise or even precise: siblings of \segDestination\ and siblings of entities along the paths may be excluded as they do not have path from \segDestination\ or to \segSource\ in \provgraph\ (e.g., \emph{log} in $Q_1$). Second, at another extreme, a provenance subgraph induced from some paths~\cite{bertram_lang@edbt10} or all paths~\cite{provabs_pmissier@ipaw14} among vertices in \segSource$\cup$\segDestination\ will only include vertices on the paths, thus exclude other contributing ancestors for \segDestination\ (e.g., \emph{model} and \emph{solver} in $Q_1$). Moreover, quantitative techniques used in other domains other than provenance cannot be applied directly either, such as keyword search over graph data techniques~\cite{kwd_search_survey@charu_book} which also do not assume that users have full knowledge of the graph, and let users use keywords to match vertices and then induce connected subgraph among keyword vertices. However, the techniques often use tree structures (e.g., least common ancestor, Steiner tree) connecting \segSource$\cup$\segDestination\ and are not aware of provenance usages\eat{ domain knowledge}, thus cannot reflect the ancestry relationships precisely. 

\begin{figure}[t!]
\centering{
\includegraphics[width=0.4\textwidth]{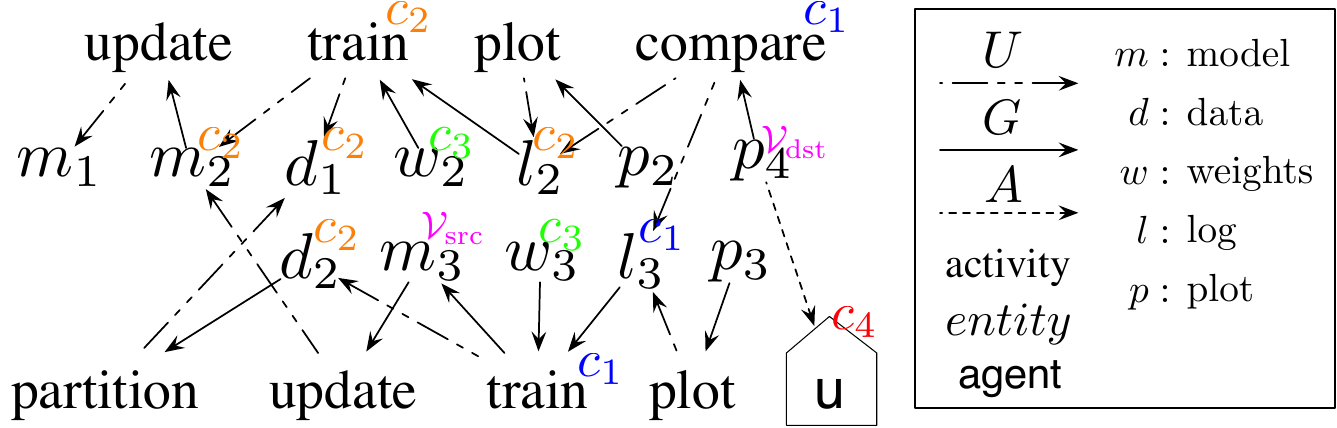}
}
\vspace{-5pt}
\caption{
{\small{In a typical repetitive model adjustment, with}} \segSource$=\{m_3\}$, \segDestination$=\{p_4\}${\small{, \opseg\ query induces }}\segInduced\ {\small{on similar adjustment paths.}}
}
\label{fig:c1_to_4_exp}
\vspace{-10pt}
\end{figure}

Instead of defining \segInduced\ quantitatively, we define \opseg\ qualitatively by a set of domain rules: \emph{a)} to be precise, \opseg\ includes other participating vertices not in the lineage and not in the paths among \segSource$\cup$\segDestination; \emph{b)} to be concise, \opseg\ utilizes the path shapes between \segSource\ and \segDestination\ given by the users as a heuristic to filter the ancestry lineage subgraph. 
We define \eat{and categorize} the rules for\eat{that generate subsets of} \segInduced\ as follows and illustrate in Fig.~\ref{fig:c1_to_4_exp}:

\subtopic{(a) Vertices on Direct Path (\segInducedOnPath)}: 
Activities and entities along any direct path \segpath{$v_j$}{$v_i$}\ between an entity $v_i\in$ \segSource\ and an entity $v_j \in$ \segDestination\ are the most important ancestry information. It helps the users answer classic provenance questions, such as reachability, i.e., whether there exists a path; workflow steps, i.e., if there is a path, what activities occurred. We refer entities and activities on such direct path as \segInducedOnPath, which is defined as\eat{ follows}: 
${\footnotesize{
\text{\segInducedOnPath} = \bigcup\limits_{v_i \in \text{\segSource}, v_j \in \text{\segDestination}}\{v_k|\ \exists_{\text{\segpath{$v_j$}{$v_i$}}}\ v_k \in \text{\segsubpath{$v_j$}{$v_i$}} \}
}}
$.
\eat{
Note not only the shortest path are of interest, but \eat{all such path }\segpath{$v_j$}{$v_i$}\ in\eat{ the DAG} \provgraph\ should be derived. }

\subtopic{(b) Vertices on Similar Path (\segInducedSimilarPath)}: 
Though \segInducedOnPath\ is important, due to the partial knowledge of the user, just considering the direct paths may miss important ancestry information including: \emph{\textbf{a)}} the entities generated together with \segDestination, \emph{\textbf{b)}} the entities used together with \segSource, and \emph{\textbf{c)}} more importantly, other entities and activities which are not on the direct path, but contribute to the derivations. 
The contributing vertices are particularly relevant to the query in our context, because data science project consists of many back-and-forth repetitive and similar steps, such as dataset splits in cross-validation, similar experiments with different hyperparameters and model adjustments in a plot (Fig.~\ref{fig:c1_to_4_exp}).\eat{configurations of preparing data in alternative ways, adjusting model templates, and evaluating experiments. }

To define the induction scope, on one hand, all ancestors w.r.t. \segDestination\ in the lineage subgraph would be returned, however it is very verbose and not concise to interpret. 
On the other hand, it is also difficult to let the user specify all the details of what should/should not be returned. 
Here we use a \ul{heuristic}: \emph{induce ancestors which are not on the direct path but contribute to \segDestination\ in a similar way, i.e., path labels from \segInducedSimilarPath\ to \segDestination\ are the same with some directed path from \segSource.} In other words, one can think it is similar to a radius concept~\cite{prov_segmentation@tapp16} to slice the ancestry subgraph w.r.t. \segDestination, but the radius is not measured by how many hops away from \segDestination\ but by path patterns between both \segDestination\ and \segSource\ \eat{entities }that are specified by the user query. 
Next we first formulate the path pattern in a context free language~\cite{intr_automata@3ed}, \cfglanguage{\similarPathPatternRule}, then \segInducedSimilarPath\ can be defined as a $L-$constrained reachability query from \segSource\ via \segDestination\ over \provgraph, only accepting path labels in the language.

A \emph{context-free} grammar (CFG) over a provenance graph \provgraph\ and a \opseg\ query $Q$ is a 6-tuple $(\Sigma, N, P, S, \text{\provgraph}, Q)$, where $\Sigma=\{$\entity,\activity,\agent$\}\cup\{$\used,\wasGeneratedBy,\wasAssociatedWith,\wasAttributedTo,\wasDerivedFrom$\}\cup$\segDestination\ is the alphabet consisting of vertex labels, edge labels in \provgraph\ and \segDestination\ vertex identifiers (e.g., {\tt id} in Neo4j) in $Q$, $N$ is a set of non-terminals, $P$ is the set of production rules, and $S$ is the start symbol. 
Each production rule in the form of $l \rightarrow\ (\Sigma\cup N)*$ defines an acceptable way of concatenations of the RHS words for the LHS non-terminal $l$. Given a CFG and a non-terminal $l_i \in N$ as the start symbol, a \emph{context-free} language (CFL), \cfglanguage{$l_i$}, is defined as the set of all finite words over $\Sigma$ by applying its production rules. 

The following CFG defines a language \cfglanguage{\similarPathPatternRule} that describes the heuristic path segment pattern for \segInducedSimilarPath\eat{ the induced vertex set}. The production rules expand \eat{inversely} from some $v_j\in$\segDestination\  both ways to reach $v_i$ and $v_k$, such that the concatenated path \segpath{$v_i$}{$v_k$} has the destination $v_j$ in the middle. 
\vspace{-3pt}
\begin{equation*}
{\small{
\begin{aligned}
  \text{\similarPathPatternRule} &\rightarrow\quad 
      \text{\wasGeneratedByInv\entity}\ \text{\similarPathPatternRule}\ \text{\entity\wasGeneratedBy} \\
      &\quad|\quad \text{\usedInv\activity}\ \text{\similarPathPatternRule}\ \text{\activity\used} \\
      &\quad|\quad \text{\wasGeneratedByInv}v_j\text{\wasGeneratedBy} & \forall v_j \in \text{\segDestination} \\
\end{aligned}
}}
\vspace{-3pt}
\end{equation*}
Now we can use \cfglanguage{\similarPathPatternRule} to define \segInducedSimilarPath\ as\eat{accordingly}: for any vertex $v_k$ in \segInducedSimilarPath, 
there should be at least a path from a $v_i \in \text{\segSource}$ going through a $v_j \in \text{\segDestination}$ \eat{and $v_k$ }then reaching some vertex $v_t$, \eat{such that}s.t. the path segment label $\text{\pathlabel}(\text{\segsubpath{$v_i$}{$v_t$}})$ is a word in $\text{\cfglanguage{\similarPathPatternRule}}$:
$
{\footnotesize{
\text{\segInducedSimilarPath}=\bigcup\limits_{v_i \in \text{\segSource}}\{v_k|\ \exists_{\text{\segpath{$v_i$}{$v_t$}}}\ \text{\pathlabel}(\text{\segsubpath{$v_i$}{$v_t$}}) \in \text{\cfglanguage{\similarPathPatternRule}} \wedge  v_k \in \text{\segpath{$v_i$}{$v_t$}}  \}
}}
$

Using CFG allows us to express the heuristic properly. Note that \cfglanguage{\similarPathPatternRule} cannot be described by regular expressions over the path(segment) label, as it can be viewed as a palindrome language~\cite{intr_automata@3ed}. 
Moreover, it allows us to extend the query easily by using other label functions, \eat{for example,}e.g., instead of \vlabelfunc$(v)$ and \elabelfunc$(e)$ whose domains are \prov\ types, using property value \vpropfunc$(v, p_i)$ or \epropfunc$(e, p_j)$ in \provgraph\ allows us to describe interesting constraints, e.g., the induced path \eat{from \segInducedSimilarPath\ to \segDestination\  }should use the same commands as the path from \segSource\ to \segDestination, or the matched entities on both sides of the path should be attributed to the same agent. For example, the former case can simply modify the second production rule in the CFG as\eat{ follows}:
%
\vspace{-3pt}
\begin{equation*}
{\small{
\begin{aligned}
      &\text{\usedInv}\ \text{\vpropfunc}(a_i, p_0)\ \text{\similarPathPatternRule}\ \text{\vpropfunc}(a_j, p_0)\ \text{\used} &\\
      &\text{s.t.}\ a_i,a_j \in \text{\activity} \wedge p_0 = \text{`command'} \wedge \text{\vpropfunc}(a_i, p_0)\ = \text{\vpropfunc}(a_j, p_0) \\
\end{aligned}
}}
\vspace{-3pt}
\end{equation*}
This is a powerful generalization that\eat{ allows \opseg\ to }\eat{constrain induction scope by describing} can describe repetitiveness and similarily ancestry paths at a very fine granularity.

\eat{
Note that regular pattern queries (RPQ) with possible path variables are not supported well in modern graph query languages and graph database~\cite{survey_graphquery@sigmodrecord12,pathquerylang_libkin@tods12,pgql_oracle@grades16}; \opseg\ cannot be evaluated directly via query facilities provided by the graph database backend. 
We develop efficient \opseg\ evaluation technique that is suitable for our needs on provenance graphs in Sec.~\ref{subsubsec:seg_qalgorithm}.
}

\subtopic{(c) Entities Generated By Activities on Path (\segInducedGenTogether)}:
As mentioned earlier, the sibling entities generated together with \segDestination\ may not be induced from directed paths. The same applies to the siblings of entities induced in \segInducedOnPath\ and \segInducedSimilarPath\eat{, if the siblings do not have paths to \segDestination}. We refer to those entities as \segInducedGenTogether\ and define it as:
$
{\footnotesize{
\text{\segInducedGenTogether} = \bigcup\limits_{v_i \in \text{\activity}\cap(\text{\segInducedOnPath}\cup\text{\segInducedSimilarPath})}\{v_{\epsilon}|\ (v_{\epsilon}, v_i) \in \text{\wasGeneratedBy} \wedge v_{\epsilon} \notin \text{\segInducedOnPath}\cup\text{\segInducedSimilarPath} \} 
}}
$

\subtopic{(d) Involved Agents (\segInducedAgents)}:
Finally, the agents \eat{in the provenance graph }may be important in some situations, e.g., from the derivation, identify who makes a mistake, like \cmd{git blame} in version control settings. On a provenance graph\eat{ \provgraph}, agents can be derived easily:
${\footnotesize{\text{\segInducedAgents} = \bigcup\limits_{v_i \in V'}\{v_u|\ v_u\in \text{\agent} \wedge (v_i, v_u) \in \text{\wasAssociatedWith} \cup \text{\wasAttributedTo} \}
}}$, where $V'$ is the union of all query vertices and other induced vertices.

\subsubsection{Boundary Criteria \boundaryCriteria} 
On the induced subgraph, \eat{besides path shapes, }the segmentation operator should be able to express users' logical boundaries when asking the ancestry queries. It is particulary useful in an interactive setting once the user examines the returned induced subgraph and wants to make adjustments. We categorize the boundary criteria support as \emph{\textbf{a)}} exclusion constraints and \emph{\textbf{b)}} expansion specifications.

First, boundaries would be constraints to exclude some parts of the graph, such as limiting\eat{processions} ownership (authorship) (\textbf{who}), time intervals (\textbf{when}), project steps (particular version, file path patterns) (\textbf{where}), understanding capabilities (neighborhood size) (\textbf{what}), etc. 
Most of the boundaries can be defined as boolean functions mapping from a vertex or edge to true or false, i.e., $b_v(v): \text{\vertexset} \mapsto \{0,1\}$, $b_e(e): \text{\edgeset} \mapsto \{0,1\}$, which can be incorporated easily to the CFG framework for subgraph induction. We define the exclusion boundary criteria as two sets of boolean functions ($\text{\boundaryCriteria}_v$ for vertices and $\text{\boundaryCriteria}_e$ for edges), which could be provided by the system or defined by the user. Then the labeling function used for defining \segInduced\  would be adjusted by applying the boundary criteria as follows:
\vspace{-3pt}
\begin{equation*}
{\small{
\text{\actualLabel}_v = \begin{cases}
\text{\vlabelfunc}(v) &  \bigwedge_{b_i\in\text{\boundaryCriteria}_v} b_i(v) = 1 \\
\varepsilon    &  \text{otherwise} \\
\end{cases}
\text{\actualLabel}_e = \begin{cases}
\text{\elabelfunc}(e) &  \bigwedge_{b_i\in\text{\boundaryCriteria}_e} b_i(e) = 1 \\
\varepsilon    &  \text{otherwise} \\
\end{cases}
}}
\vspace{-3pt}
\end{equation*}

\noindent In other words, a vertex or an edge that satisfies all exclusion boundary conditions, is mapped to its original label. Otherwise the empty word ($\varepsilon$) is used as its label, so that 
paths having that vertex will not satisfy \cfglanguage{\similarPathPatternRule}.

Second, instead of exclusion constraints, the user may wish to expand the induced subgraph. We allow the users to specify expansion criteria, $\text{\boundaryCriteria}_x = \{b_x(V_x, k)\}$, denoting including paths \eat{from entities} which are $k$ activities away from entities in $V_x \subseteq$ \segInduced. 
\eat{For example, }In Fig.~\ref{fig:example_query_segmentation}, $Q_1$ excludes $\{$\wasAttributedTo, \wasDerivedFrom$\}$ edges via $\text{\boundaryCriteria}_e$ and expands by $\text{\boundaryCriteria}_x = \{b_x(\text{\emph{weight-v2}}, 2)\}$, so $\{$\emph{update-v2}, \emph{model-v1}$\}$ are included.

\subsection{Query Evaluation}
\label{subsubsec:seg_qalgorithm}

\subsubsection{Overview: Two-Step Approach}
Given a \opseg\eat{(\segSource,\segDestination,\boundaryCriteria)} query, we separate the query evaluation into two steps: 1) \textbf{induce}: induce \segInduced\ and construct the induced graph \segsubgraph\ using \segSource and \segDestination\eat{in memory}, 2) \textbf{adjust}: apply \boundaryCriteria\ interactively to filter induced vertices or retrieve more vertices from the property graph store backend. The rationale of the two-step approach is that the operator is designed for the users with partial knowledge who are willing to understand a local neighborhood in the provenance graph. Any induction heuristic applied would be unlikely to match the user's implicit interests and would require back-and-forth explorations. 

In the rest of the discussion, we assume \textbf{\emph{a)}} the provenance graph is stored in a backend property graph store, with constant time complexity to access arbitrary vertex and arbitrary edge by corresponding primary identifier; \textbf{\emph{b)}} given a vertex, both its incoming and outgoing edges can be accessed equally, with linear time complexity w.r.t. the in- or out-degree.
In our implementation\eat{ (Sec.~\ref{sec:system})}, \eat{we use }Neo4j\eat{ as our storage backend, which} satisfies the conditions -- both nodes and edges are accessed via their {\tt id}. 

\begin{figure}[b!]
\vspace{-5.0pt}
\begin{lstlisting}[caption={{\footnotesize{Cypher $Q_1$ for}} \cfglanguage{\similarPathPatternRule},\segSource{\small{$=\{\alpha_1,\alpha_2\}$}},\segDestination{\small{$=\{\beta_1,\beta_2\}$}}}, label=cypher:q0]
 match p1=(b:E)<-[:U|G*]-(e1:E) with p1 
 where id(b) in [%*$\alpha_1$, $\alpha_2$*)] and id(e1) in [%*$\beta_1$, $\beta_2$*)] 
 match p2=(c:E)<-[:U|G*]-(e2:E) 
 where id(e2) in [%*$\beta_1$, $\beta_2$*)] and 
   extract(x in nodes(p1) | labels(x)[0]) 
     = extract(x in nodes(p2) | labels(x)[0]) and 
   extract(x in relationships(p1) | type(x)) 
     = extract(x in relationships(p2) | type(x)) 
 return p2; 
\end{lstlisting}
\end{figure}

\subsubsection{Induce Step}
Given \segSource\ and \segDestination, \opseg\ induces \segInduced\ \eat{which consists of}with four categories. We mainly focus our discussion on inducing vertices on direct and similar paths, as the other two types\eat{, i.e., sibling entities and related agents} can be derived in a straightforward manner by scanning 1-hop neighborhoods of the first two sets of results.

\topic{Cypher}: 
The definition of vertices on similar path requires a context-free language, and cannot be expressed by a regular language. When developing the system, we realize it can be decomposed into two regular language path segments, and express the query using path variables~\cite{survey_graphquery@sigmodrecord12,pathquerylang_libkin@tods12}. We handcraft a Cypher query shown in Query~\ref{cypher:q0}. The query uses \segSource\ (\cmd{b}) and \segDestination\ (\cmd{e1}) to return all directed paths \segInducedOnPath\ via path variables (\cmd{p1}), and uses Cypher \cmd{with} clause to hold the results. The second \cmd{match} finds the other half side of the \similarPathPatternRule\ via path variable \cmd{p2} which then joins with \cmd{p1} to compare the node-by-node and edge-by-edge conditions to induce \segInducedSimilarPath. If we do not need to check properties, then we can use \cmd{length(p1) = length(p2)} instead of the two \cmd{extract} clauses.
However, as shown later in the evaluation (Sec.~\ref{sec:exp}), Neo4j takes more than 12 hours to return results for even very small graphs with about a hundred vertices. Note that \eat{regular pattern queries (}RPQ\eat{)} with path variables are not supported well in modern graph query languages and graph database~\cite{survey_graphquery@sigmodrecord12,pgql_oracle@grades16}, we develop our own \opseg\ algorithm for provenance graphs.

\topic{CFL-reachability}: 
Given a vertex $v$ and a CFL $L$, the problem of finding all vertieces $\{u\}$ such that there is a path \segpath{$v$}{$u$} with label $\text{\pathlabel}(\text{\segpath{$v$}{$u$}}) \in L$ is often referred as \emph{single source} \emph{CFL-reachability} (CFLR) problem or \emph{single source} \emph{L-Transitive Closure} problem~\cite{cflr_to_datalog@pods90,reps_pa_reachability@islp97}. The \emph{all-pairs} version, which aims to find all such pairs of vertices connected by a $L$ path of the problem, has the same complexity. 
As \segSource\ would be all vertices, we do not distinguish between the two in the rest of the discussion.
Though the problem has been first studied in our community~\cite{cflr_to_datalog@pods90}, there is little follow up and support in the context of modern graph databases (Sec.~\ref{sec:related_work}). CFLR finds its main application in programming languages and is recognized as a general formulation for many program analysis tasks~\cite{reps_pa_reachability@islp97}. \eat{On graph representations of programs, program analysis tasks such as program slicing and pointer analysis, can be described in a CFL to specify path patterns (Sec.~\ref{sec:related_work}). }

State of the art CFLR algorithm~\cite{chaudhuri_subcublic@popl08} solves the problem in $O(n^3/log(n))$ time and $O(n^2)$ space w.r.t. the number of vertices in the graphs. It is based on a classic cubic time dynamic programming scheme~\cite{reps_pa_reachability@islp97,reps_cflr_baseline@tcs00} which derives production facts non-repetitively via graph traversal, and uses the method of four Russians~\cite{fourrussian70} during the traversal. In the rest of the paper, we refer it as \baselineAlg. We analyze it on provenance graphs for \cfglanguage{\similarPathPatternRule}, then present improvement techniques. The details of \baselineAlg\ and proofs are included in Appendix.

\eat{
\begin{figure}[t!]
\begin{equation*}
{\small{
\begin{aligned}
  r_0: \text{\sc Qd} &\rightarrow\quad v_j                                        & \forall v_j \in &\text{\segDestination}                                      & \\
  r_1: \text{\sc Lg} &\rightarrow\quad \text{\wasGeneratedByInv}\ \text{\sc Qd}   & r_3: \text{\sc La} &\rightarrow\quad \text{\activity}\ \text{\sc Rg}  & r_6: \text{\sc Ru} &\rightarrow\quad \text{\sc Lu}\ \text{\used}    \\
   &\quad|\quad  \text{\wasGeneratedByInv}\ \text{\sc Re}       & r_4: \text{\sc Ra} &\rightarrow\quad \text{\sc La}\ \text{\activity}  & r_7: \text{\sc Le} &\rightarrow\quad \text{\entity}\ \text{\sc Ru}  \\
  r_2: \text{\sc Rg} &\rightarrow\quad \text{\sc Lg}\ \text{\wasGeneratedBy}      & r_5: \text{\sc Lu} &\rightarrow\quad \text{\usedInv}\ \text{\sc Ra}   & r_8: \text{\sc Re} &\rightarrow\quad \text{\sc Le}\ \text{\entity}  \\
\end{aligned}
}}
\end{equation*}
\vspace{-15pt}
\caption{\similarPathPatternRule\ Normal Form\eat{, \similarPathPatternRule\ $\rightarrow \text{\sc Re}$}. %
$\text{\sc Lg} \subseteq \text{\activity}\times\text{\entity}$; %
$\text{\sc Rg, La, Ra} \subseteq \text{\activity}\times\text{\activity}$; %
$\text{\sc Lu} \subseteq \text{\entity}\times\text{\activity}$; %
$\text{\sc Ru, Le, Re, Qd} \subseteq \text{\entity}\times\text{\entity}$. %
}
\label{fig:simprov_normalized}
\vspace{-16pt}
\end{figure}
}

\eat{
We first describe the algorithm briefly and then present improvement techniques for \cfglanguage{\similarPathPatternRule} on provenance graphs.
}

Given a CFG, \baselineAlg\ works on its normal form~\cite{intr_automata@3ed}, where each production has at most two RHS symbols, i.e., $N \rightarrow AB$ or $N \rightarrow A$.  
The normal form of \similarPathPatternRule\ is listed in Fig.~\ref{fig:simprov_normalized}.
At a high level, the algorithm traverses the graph and uses grammar as a guide to find new production facts $N(i,j)$, where $N$ is a LHS nonterminal, $i,j$ are graph vertices, and the found fact $N(i,j)$ denotes that there is a path from $i$ to $j$ whose path label satisfies $N$. 
To elaborate, similar to BFS, it uses a worklist $W$ (queue) to track newly found fact $N(i,j)$ and a \emph{fast set} data structure $H$ with time complexity $O(n/log(n))$ for set diff/union and $O(1)$ for insert to memorize found facts. 
\eat{In the beginning, all facts $F(i,j)$ from all single RHS symbol rules $F \rightarrow A$ are enqueued. In \similarPathPatternRule\ case ($r_0$ in Fig.~\ref{fig:simprov_normalized}), each $v_j \in$\segDestination\ is added to $W$ as $\text{\sc Qd}(v_j, v_j)$. From $W$, it\eat{the algorithm} processes one fact $F(i,j)$ at a time until $W$ is empty. When processing a dequeued fact $F(i,j)$, if $F$ appears in any rule in \eat{the following }cases:
$
{\footnotesize{
N(i,j)\rightarrow F(i,j); N(i,v)\rightarrow F(i,j)A(j,v); N(u,j)\rightarrow A(u,i)F(i,j)
}}
$,\\
the new LHS fact $N(i,v)$ is derived by set diff $\{v \in A(j,v)\}\setminus \{v \in N(i,v)\}$ or $N(u,j)$ by $\{u \in A(u,i)\}\setminus \{u \in N(u,j)\}$ in $H$. }
Then the new facts of $N$ are added to $H$ to avoid repetition and $W$ to explore it later. Once $W$ is empty, the start symbol $L$ facts $L(i,j)$ in $H$ include all vertices pairs $(i,j)$ which have a path with label that satisfies $L$. 
If path is needed, a parent table would be used similar to BFS. 
In \similarPathPatternRule\ (Fig.~\ref{fig:simprov_normalized}), the start symbol is {\sc Re}, $\forall v_i \in \text{\segSource}$, $\text{\sc Re}(v_i, v_t)$ facts include all $v_t$, s.t. between them there is $\text{\pathlabel}({\text{\segsubpath{i}{t}}}) \in \text{\cfglanguage{\similarPathPatternRule}}$. 

\topic{\cfglanguage{\similarPathPatternRule}-reachability on \prov}: 
\eat{
Next we study the performance of \baselineAlg\ for \similarPathPatternRule\ on a \prov\ graph, and show the \emph{fast set} method is not suitable for \prov\ graph. Then we further explore grammar and \prov\ graph properties, instead of normal form, we rewrite the grammar to allow several pruning strategies and propose a linear-time algorithm if $|\text{\segDestination}|$ can be viewed as a constant.
}
In our context, the provenance graph is often sparse, and both the numbers of entities that an activity uses and generates can be viewed as a small constant, however the domain size of activities and entities \eat{\activity\ and \entity\ }are potentially large. The following lemma shows show the \emph{fast set} method is not suitable for \prov\ graph.

\begin{lemma}
\baselineAlg\ solves \cfglanguage{\similarPathPatternRule}-reachability on a \prov\ graph in 
$O({|\text{\activity}||\text{\entity}|^2}/{\log|\text{\activity}|} + {|\text{\entity}||\text{\activity}|^2}/{\log|\text{\entity}|})$
time if using \emph{fast set}.
Otherwise, it solves it in $O(|\text{\wasGeneratedBy}||\text{\entity}| + |\text{\used}||\text{\activity}|)$ time.
\end{lemma}
\eat{
\begin{proof}
On \similarPathPatternRule\ normal form (Fig.~\ref{fig:simprov_normalized}), for $i \in [1,8]$, \baselineAlg\ derives $r_i$ LHS facts by a $r_{i-1}$ LHS fact dequeued from $W$ (Note it also derives $r_1$ from $r_8$). For $i \in \{1,2\}$, $r_i(u,v)$ uses \wasGeneratedBy\ edges in the graph during the derivation, e.g., from $r_8$ LHS $\text{\sc Re}$ to $r_1: \text{\sc Lg}(u,v) \rightarrow \text{\wasGeneratedByInv}(u, k)\ \text{\sc Re}(k, v)$. As $\text{\sc Re}(k,v)$ can only be in the worklist $W$ once, we can see that each 3-tuple $(u,k,v)$ is formed only once on the RHS and there are at most $|\text{\wasGeneratedBy}||\text{\entity}|$ of such 3-tuples. To make sure $\text{\sc Lg}(u,v)$ is not found before, $H$ is checked. If not using fast set but a $O(1)$ time procedure for each instance $(u,k,v)$, then it takes $O(|\text{\wasGeneratedBy}||\text{\entity}|)$ to produce the LHS; on the other hand, if using a \emph{fast set} on $u's$ domain \activity\ for each $u$, for each $\text{\sc Re}(k,v)$, $O({|\text{\activity}|}/{\log|\text{\activity}|})$ time is required, thus it takes $O({|\text{\activity}||\text{\entity}|^2}/{\log|\text{\activity}|})$ in total. Applying similar analysis on $r_5$ and $r_6$ using \used\ to derive new facts, we can see it takes $O({|\text{\entity}||\text{\activity}|^2}/{\log|\text{\entity}|})$ with fast set and $O(|\text{\used}||\text{\activity}|)$ without fast set. Finally $r_3, r_4$ and $r_7, r_8$ can be viewed as following a vertex self-loop edge and do not affect the complexity result. 
\vspace{-3pt}
\end{proof}
}

\eat{
In our context, the \prov\ graph is often sparse, and both the numbers of entities that an activity uses and generates can be viewed as a small constant, however the domain size of \activity\ and \entity\ are potentially large. }
The lemma also reveals a quadratic time scheme for \\\cfglanguage{\similarPathPatternRule}-reachability if we can view the average in-/out- degree as a constant. Note that the quadratic time complexity is not surprising, as \similarPathPatternRule\ is a linear CFG, i.e., there is at most one nonterminal on RHS of each production rule. The CFLR time complexity for any linear grammar on general graphs $G(V,E)$ have been shown in theory as $O(|V||E|)$ by a transformation to general transitive closures~\cite{cflr_to_datalog@pods90}.

\begin{figure}[t]
\begin{equation*}
{\small{
\begin{aligned}
r'_1: \text{\sc Ee} &\rightarrow\quad v_j\quad\quad  \forall v_j \in \text{\segDestination} &\quad r'_2: \text{\sc Aa} &\rightarrow\quad \text{\wasGeneratedByInv}\ \text{\sc Ee}\ \text{\wasGeneratedBy}       \\
                    &\quad|\quad  \text{\usedInv}\ \text{\sc Aa}\ \text{\used}   & &\quad|\quad  \text{\activity}\ \text{\sc Aa}\ \text{\activity} \\
                    &\quad|\quad  \text{\entity}\ \text{\sc Ee}\ \text{\entity}   \\
\end{aligned}
}}
\end{equation*}
\vspace{-15pt}
\caption{\eat{Proposed }\similarPathPatternRule\ Rewriting, \similarPathPatternRule\ $\rightarrow \text{\sc Ee}$. %
$\text{\sc Aa} \subseteq \text{\activity}\times\text{\activity}$; %
$\text{\sc Ee} \subseteq \text{\entity}\times\text{\entity}$. %
}
\label{fig:simprov_new_normal_form}
\vspace{-12pt}
\end{figure}

\topic{Rewriting \similarPathPatternRule}: 
Most CFLR algorithms require the normal form mentioned earlier. However, under the normal form, it \textbf{a)} introduces more worklist entries, and \textbf{b)} misses important grammar properties. Instead, we rewrite \similarPathPatternRule\ as shown in Fig.~\ref{fig:simprov_new_normal_form}, and propose \opsegAlg\ and \opsegAlgTst\ by adjusting \baselineAlg. Comparing with standard normal forms, $r'_1$ and $r'_2$ have more than two RHS symbols. \opsegAlg\ utilizes the rewritten grammar and \prov\ graph properties to improve \baselineAlg. Moreover, \opsegAlgTst\ solves \cfglanguage{\similarPathPatternRule}-reachability on a \prov\ graph in linear time and sublinear space if viewing $|\text{\segDestination}|$ as constant. The properties of the rewritten grammar and how \eat{\opsegAlg\ and \opsegAlgTst\ }to utilize them are described below, which can be used in other CFLR problems:

\textbf{\emph{a)}} \ul{Reduction for Worklist tuples}: 
Note that $r'_2$ in Fig.~\ref{fig:simprov_new_normal_form}, 
\eat{\\}
$
{\footnotesize{
\text{\sc Aa}(a_1,a_2) \rightarrow \text{\wasGeneratedByInv}(a_1, e_1)\ \text{\sc Ee}(e_1, e_2)\ \text{\wasGeneratedBy}(e_2, a_2)
}}
$, combines rules $r_1, r_2$ in the normal form, i.e., ${\footnotesize{\text{\sc Rg}(a_1,a_2) \rightarrow \text{\sc Lg}(a_1, e_2) \text{\wasGeneratedBy}(e_2, a_2)}}$ and ${\footnotesize{\text{\sc Lg}(a_1,e_2) \rightarrow \text{\wasGeneratedByInv}(a_1, e_1) \text{\sc Re}(e_1, e_2)}}$.
\eat{
combines rules $r_1, r_2$ in the normal form, i.e., ${\footnotesize{\text{\sc Rg}(a_1,a_2) \rightarrow \text{\sc Lg}(a_1, e_2) \text{\wasGeneratedBy}(e_2, a_2)}}$ and ${\footnotesize{\text{\sc Lg}(a_1,e_2) \rightarrow \text{\wasGeneratedByInv}(a_1, e_1) \text{\sc Re}(e_1, e_2)}}$, is derived by:
\eat{\\}
$
{\footnotesize{\quad
\text{\sc Aa}(a_1,a_2) \rightarrow \text{\wasGeneratedByInv}(a_1, e_1)\ \text{\sc Ee}(e_1, e_2)\ \text{\wasGeneratedBy}(e_2, a_2)
}}
$
}
Instead of enqueue $\text{\sc Lg}(a_1, e_2)$ and then $\text{\sc Rg}(a_1, a_2)$, \opsegAlg\ adds $\text{\sc Aa}(a_1, a_2)$ to $W$ directly. 
In the \eat{previous }normal form, there may be other cases that can also derive $\text{\sc Aa}(a_1, a_2)$, i.e., in presence of $\text{\sc Lg}(a_1, e_k)$ and $\text{\wasGeneratedBy}(e_k, a_2)$. In the worst case, \baselineAlg\ enqueued $|$\entity$|$ number of {\sc Lg} in $W$ which later find the same fact $\text{\sc Rg}(a_1, a_2)$. It's worth mentioning that in \opsegAlg\, because $\text{\sc Aa}(a_1, a_2)$ now would be derived by many $\text{\sc Ee}(e_i, e_j)$ in $r'_2$, before adding it to $W$, we need to check if it is already in $W$\eat{. In \opsegAlg}. We use two pairs of bitmaps for {\sc Ee} and {\sc Aa} for $W$ and $H$\eat{ respectively}, the space cost is 
$O({|\text{\entity}|^2}/{\log|\text{\entity}|} + {|\text{\activity}|^2}/{\log|\text{\activity}|})$. 
Compressed bitmaps would be used to improve space usage at the cost of non-constant time random read/write.

\textbf{\emph{b)}} \ul{Symmetric property}: 
In the rewritten grammar, both nonterminals {\sc Ee} and {\sc Aa} are symmetric, i.e., {\sc Ee}$(e_i, e_j)$ implies {\sc Ee}$(e_j, e_i)$, {\sc Aa}$(a_i, a_j)$ implies {\sc Aa}$(a_j, a_i)$, which is not held in normal forms. Intuitively {\sc Ee}$(e_1, e_2)$ means some path label from $e_1$ to $v_j \in$\segDestination\ is the same with some path label from $e_2$ to $v_j$. Using symmetric property, in \opsegAlg, we can use a straightforward \textbf{\emph{pruning strategy}}: only process $(e_i, e_j)$ in both $H$ and $W$ if {\tt id}$(e_i) \leq${\tt id}$(e_j)$, and $(a_i, a_j)$ if {\tt id}$(a_i) \leq${\tt id}$(a_j)$; and an \textbf{\emph{early stopping rule}}: for any {\sc Aa}$(a_i, a_j)$ that both $a_i$'s and $a_j$'s order of being is before all \opseg\ query \segSource\ entities, we do not need to proceed further. Note the early stopping rule is \similarPathPatternRule\ and \prov\ graph specific, while solving general CFLR, even in the \emph{single-source} version, cannot take source information and we need to evaluate until the end. Though both strategies \eat{used by \opsegAlg\ }do not improve the worst-case time complexity, they are very useful in realistic \prov\ graphs (Sec.~\ref{sec:exp}).

\textbf{\emph{c)}} \ul{Transitive property}: 
By definition \similarPathPatternRule\ does not have transitivity, i.e., given {\sc Ee}$(e_1, e2)$ and {\sc Ee}$(e_2, e_3)$, it does not imply {\sc Ee}$(e_1, e_3)$. This is because a \opseg\ query allows multiple \segDestination, {\sc Ee}$(e_1, e_2)$ and {\sc Ee}$(e_2, e_3)$ may be found due to different $v_j \in$\segDestination. However, if we evaluate $v_j \in$\segDestination\ separately, then {\sc Ee} and {\sc Aa} have transitivity, which leads to a linear algorithm \opsegAlgTst\ for each $v_j$: instead of maintaining {\sc Ee}$(e_i, e_j)$ or {\sc Aa}$(a_i, a_j)$ tuples in $H$ and $W$, we can use a set $[e]_m$ or $[a]_n$ to represent an equivalence class at iteration $m$ or $n$ where any pair in the set is a fact of {\sc Ee} or {\sc Aa} respectively.  If at iteration $m$, the current $W$ holds a set $[e]_m$, then $r'_2:$ {\sc Aa}$(a,a) \rightarrow$\wasGeneratedByInv$(a,e)${\sc Ee}$(e,e)$\wasGeneratedBy$(e,a)$ is used to infer the next $W$ (a set $[a]_{m+1}$); otherwise, $W$ must hold a set $[a]_m$, then \eat{{\sc Ee}$(e,e) \rightarrow$\usedInv$(e,a)${\sc Aa}$(a,a)$}similarly $r'_1$ is used to infer next equivalence class $[e]_{m+1}$ as the next $W$. In the first case, as there are at most $|$\wasGeneratedBy$|$ possible $(a,e)$ tuples, the step takes $O(|\text{\wasGeneratedBy}|)$ time; in the later case, similarly the step takes $O(|\text{\used}|)$ time. 
The algorithm returns vertices in any equivalence classes $[v_i]_m, \text{s.t.} v_i \in \text{\segSource}$. Overall, because there are multiple \segDestination\ vertices, the algorithm runs in $O(|\text{\segDestination}||\text{\wasGeneratedBy}| + |\text{\segDestination}||\text{\used}|)$ time and $O({|\text{\entity}|}/{log|\text{\entity}|} + {|\text{\activity}|}/{log|\text{\activity}|})$ space. The early stop rule can be applied\eat{ as well}, instead of a pair of activities\eat{ in \opsegAlg}, in \opsegAlgTst\ all activities in an equivalent class $[a]_{m}$ are compared with entities in \segSource\ in terms of the order of being; while the pruning strategy is not necessary, as all pairs are represented compactly in an equivalent class.

\begin{theorem}
\opsegAlgTst\ solves \cfglanguage{\similarPathPatternRule}-reachability \eat{on a \prov\ graph }in $O(|\text{\wasGeneratedBy}| + |\text{\used}|)$ time\eat{and $O(|\text{\entity}|/\log|\text{\entity}| + |\text{\activity}|/log|\text{\activity}|)$ space}, if viewing $|\text{\segDestination}|$ \eat{can be viewed}as a constant.
\end{theorem}

\subsubsection{Adjust Step}
Once the induced graph \segsubgraphFull\ is derived, the adjustment step applies boundary criteria to filter existing vertices and retrieve more vertices. Comparing with induction step, applying boundary criteria is rather straightforward. For exclusion constraints $\text{\boundaryCriteria}_v$ and $\text{\boundaryCriteria}_e$, we apply them on vertices and edges in \segsubgraphFull\ linearly if present. For $\text{\boundaryCriteria}_x$, we traverse the backend store with specified entities for $2k$ hops through \wasGeneratedByInv and \usedInv edges to their ancestry activities and entities. To support back and forth interaction, we cache the induced graph instead of inducing multiple times. We expect $k$ is small constant in our context as the generated graph is for humans to interpret, otherwise, a reachability index is needed. 
\eat{
For other purposes where the two-step approaches are not ideal, the exclusion constraints $\text{\boundaryCriteria}_v$ and $\text{\boundaryCriteria}_e$, and expansion criteria $\text{\boundaryCriteria}_x$ can be evaluated together using \baselineAlg, \opsegAlg\ and \opsegAlgTst\ with small modifications on the grammar. In \baselineAlg\, the label function $\mathcal{F}_v$ of $\text{\boundaryCriteria}_v$ can be applied at $r_0, r_3, r_4, r_7, r_8$ on \activity\ or \entity, while $\mathcal{F}_e$ of $\text{\boundaryCriteria}_e$ can be applied at rest of the rules involving \used\ and \wasGeneratedBy. For \opsegAlg\ and \opsegAlgTst, $\mathcal{F}_v$ and $\mathcal{F}_e$ can be applied together at $r'_1, r'_2$.}

\eat{
\subsubsection{Discussion}
We mainly focus on developing ad-hoc query evaluation schemes. As of now, the granularity of provenance in our context is at the level of commands executions, the number of activities are constrained by project members' work rate. In case when the \prov\ graph becomes extremely large, indexing techniques and incremental algorithms are more practical. We leave them as future steps.
}


\section{Summarization Operation}
\label{subsec:sum_op}

In a collaborative analytics project, 
collected provenance graph of the repetitive modeling trails  
reflects different roles and work routines of the team members and 
records steps of various pipelines, some of which having subtle differences.
Using \opseg, the users can navigate to their segments of interest, which 
may be about similar pipeline steps.
\eat{
For example, the query result of a single \opseg$($\segSource,\segDestination,\boundaryCriteria$)$, e.g., `\emph{yesterday's input data and prediction result}', shows a pipeline subgraph $\text{\segsubgraph}_1$ about how \segDestination\ (\emph{prediction result}) was derived from \segSource\ (\emph{input data}) together with other induced \segInduced\ (e.g., modeling steps). 
As there is no skeleton for the pipeline, given a different tuple $({\mathcal V^2_{src}},{\mathcal V^2_{dst}},\text{\boundaryCriteria}^2)$ to get another segment $\text{\segsubgraph}_2$, e.g., `\emph{today's input data and model result}', the pipeline subgraph $\text{\segsubgraph}_2$ may or may not be the same as $\text{\segsubgraph}_1$. 
} 
Given a set of segments, our design goal of \opsum\ is to produce a precise and concise provenance summary graph, \opsumgraph, which will not only allow the users to see commonality among those segments of interests (e.g., \emph{yesterday's and today's pipelines are almost the same}), but also let them understand alternative routines (e.g., \emph{an old step excluded in today's pipeline}). 
Though no workflow skeleton is defined, with that ability, \opsumgraph\ would enable the users to reason about prospective provenance in evolving workflows of analytics projects.

\subsection{Semantics of Summarization ({\opsum})}
Although there are many graph summarization schemes proposed over the years~\cite{summary_tutorial@pvldb17} (Sec.~\ref{sec:related_work}), they are neither aware of provenance domain desiderata~\cite{agg_lucmoreau@gam15} nor the meaning of \opseg\ segments. 
Given a set of segments \segmentset, each \segsubgraph$_i$ of which is a \opseg\ result, a \opsum\ query is designed to take a 3-tuple $(\text{\segmentset}, \text{\propagg}, \text{\provtypehop{k}})$ as input which describes the level of details of vertices and constrains the rigidness of the provenance; then it outputs a \eat{minimum }provenance summary graph (\opsumgraph).

\subsubsection{Property Aggregations \& Provenance Types of Vertices} 
\eat{Given a set of segments \segmentset, each \segsubgraph$_i$ of which is a \opseg\ result, t}To combine vertices and edges across the segments in \segmentset, we first introduce two concepts: \emph{\textbf{a)}} property aggregation (\propagg) and \emph{\textbf{b)}} provenance type (\provtypehop{k}), which \opsum\ takes as input and allow the user to obfuscate vertex details and retain structural constraints.

\topic{Property Aggregation (\propagg)}:
Similar to an attribute aggregation summarization query on a general graph~\cite{yytian_summary@sigmod08}, depending on the granularity level of interest, not all the details of a vertex are interesting to the user and some properties should be omitted, so that they can be combined together\eat{. For example}; e.g., in Example~\ref{exp:sumop_q3}, the user may neither care who performs an activity, nor an activity's detailed configuration; in the former case, all agent type vertices regardless of their property values (e.g., name) should be indistinguishable and in the latter case, the same activity type vertices even with different \eat{configuration settings}property values (e.g., training parameters) in various \opseg\ segments should be viewed as if the same thing\eat{ (e.g., \emph{a training activity})} has happened. 

Formally, \ul{\emph{property aggregation}} \propagg\ is a 3-tuple (\propaggtyped{\entity}, \propaggtyped{\activity}, \propaggtyped{\agent}), where each \eat{of the tuple }element is a subset of the \prov\ graph property types, i.e., \propaggtyped{\entity},\propaggtyped{\activity},\propaggtyped{\agent}$\subseteq$\property\eat{\ (Definition~\ref{def:provgraph})}. \eat{When used i}In a \opsum\ query, it discards other irrelevant properties for each vertex type, e.g., properties of entity \entity\ type in \property$\setminus$\propaggtyped{\entity} are ignored. For example, in Fig.~\ref{fig:example_query_summarization}, \propaggtyped{\entity}$=\{\text{\emph{`filename'}}\}$, \propaggtyped{\activity}$=\{\text{\emph{`command'}}\}$, \propaggtyped{\agent}$=\emptyset$, so properties such as version of the entity, details of an activity, names of the agents are ignored.

\topic{Provenance Type (\provtypehop{k})}:
In contrast with general-purpose graphs, in a provenance graph, the vertices with identical labels and property values may be very different~\cite{agg_lucmoreau@gam15}. For example, two identical activities that use different numbers of inputs or generate different entities should be considered different (e.g., \emph{update-v2} and \emph{update-v3} in Fig.~\ref{fig:example_query_segmentation}). 
In~\cite{agg_lucmoreau@gam15}, Moreau proposes to concatenate edge labels recursively\eat{using a recursive definition } over a vertex's $k$-hop neighborhood to assign a vertex type for preserving provenance meaning\eat{ later aggregation}. However, the definition ignores in-/out-degrees\eat{ of vertices,} and\eat{ the recursive definition} is exponential w.r.t. to $k$. \eat{It is worth mentioning that the former issue occurs in bisimulation-based method as well~\cite{bisimulation@icde02}. }

We extend the idea of preserving provenance meaning of a vertex and use the $k$-hop local neighborhood of a vertex to capture its \ul{\emph{provenance type}}: given a \opseg\ segment \segsubgraphFull, and a constant $k$, $k \geq 0$, provenance type \provtypehop{k}$(v)$ is a function that maps a vertex $v \in$\segv\ to its $k$-hop\eat{ neighborhood} subgraph in its segment \segsubgraph, \provtypehop{k}$\subseteq$\segsubgraph. For example, in Fig.~\ref{fig:example_query_summarization}, $k=1$, thus provenance type of vertices is the 1-hop neighborhood, vertices with label \emph{`update'}, \emph{`model'} and \emph{`solver'} all have two different provenance types (marked as `t1', `t2'). 

Note one can generalize the definition of \provtypehop{k}$(v)$ as a subgraph within $k-$hop neighborhood of $v$ satisfying a pattern matching query, which has been proposed in~\cite{wenfei_keys@pvldb15} with application to entity matching where similar to provenance graphs, just using the vertex properties are not enough to represent the conceptual identity of a vertex.

\topic{Vertex Equivalence Relation (\vertexeqrel)}:
Given \segmentset$=\{\text{\segsubgraphFullWithI}\}$, denoting the union of vertices as \segmentsetallv$=\bigcup_i\text{\segv}_i$, with the user specified property aggregation \propagg\ and provenance type \provtypehop{k}, we define a binary relation \vertexeqrel\ over \segmentsetallv$\times$\segmentsetallv, s.t.\eat{such that} for each vertex pair $(v_i, v_j)\in\text{\vertexeqrel}$: 
\begin{enumerate}[leftmargin=18pt,label=\emph{\alph*})]{}
\item 
vertex labels are the same, i.e., $\text{\vlabelfunc}(v_i) = \text{\vlabelfunc}(v_j)$;
\item
all property values in \propagg\ are equal, i.e., 
$\forall_{p\in\text{\propagg}}\ \text{\vpropfunc}(v_i, p)=\text{\vpropfunc}(v_j, p)$;
\item
\provtypehop{k}$(v_i)$ and \provtypehop{k}$(v_j)$ are graph isomorphic w.r.t. the vertex label and properties in \propagg, i.e., there is a bijection $f$ between $V_i\in$\provtypehop{k}$(v_i)$ and $V_j\in$\provtypehop{k}$(v_j)$, s.t., $f(v_m) = v_n$ if {\textbf{\small{1)}}} $\text{\vlabelfunc}(v_m) = \text{\vlabelfunc}(v_n)$, {\textbf{\small{2)}}} $\forall_{p\in\text{\propagg}}\ \text{\vpropfunc}(v_m, p)=\text{\vpropfunc}(v_n, p)$, \\{\textbf{\small{3)}}} $\forall (v_m,v_t)\in E_i, (v_n,f(v_t))\in E_j$. 
\end{enumerate}
\eat{
\noindent\textbf{a)} vertex labels are the same, i.e., $\text{\vlabelfunc}(v_i) = \text{\vlabelfunc}(v_j)$, \\
\noindent\textbf{b)} all property values in \propagg\ are equal, i.e., 
$\forall_{p\in\text{\propagg}}\ \text{\vpropfunc}(v_i, p)=\text{\vpropfunc}(v_j, p)$,\\ 
\noindent\textbf{c)} \provtypehop{k}$(v_i)$ and \provtypehop{k}$(v_j)$ are graph isomorphic w.r.t. the vertex label and properties in \propagg, i.e., there is a bijection $f$ between $V_i\in$\provtypehop{k}$(v_i)$ and $V_j\in$\provtypehop{k}$(v_j)$, s.t., $f(v_m) = v_n$ if 1) $\text{\vlabelfunc}(v_m) = \text{\vlabelfunc}(v_n)$, 2) $\forall_{p\in\text{\propagg}}\ \text{\vpropfunc}(v_m, p)=\text{\vpropfunc}(v_n, p)$, 3) $\forall (v_m,v_t)\in E_i, (v_n,f(v_t))\in E_j$. 
}

\noindent It is easy to see that \vertexeqrel\ is an equivalence relation on \segmentsetallv\ by inspection. Using \vertexeqrel, we can derive a partition \vertexeqrelpartition\ of \segmentsetallv, s.t., each set in the partition is an equivalence class by \vertexeqrel, denoted by $[v]$, s.t., $[v_i]\cap[v_j]=\emptyset$ and $\bigcup_i [v_i] = \text{\segmentsetallv}$. For each $[v]$, we can define its canonical label, e.g., the smallest vertex {\tt id}, for comparing vertices.

In other words, vertices in each equivalence class $[v]$ by \vertexeqrel\ describe the homogeneous candidates which can be merged by \opsum. Its definition not only allows the users to specify \emph{property aggregations} \propagg\ to obfuscate unnecessary details in different resolutions, but also allows the users to set \emph{provenance types} \provtypehop{k} \eat{in order} to preserve local structures and ensure the meaning of provenance of a merged vertex. 

\subsubsection{Provenance Summary Graph (\opsumgraph):}
Next, we define the output of \opsum, the provenance summary graph, \opsumgraph. 

\topic{Desiderata}: 
Due to the nature of provenance, the produced \opsumgraph\ should be \emph{precise}, i.e., we should preserve paths that exist in one or more segments, at the same time, we should not introduce any path that does not exist in any segment. On the other hand, \opsumgraph\ should be concise; the more vertices we can merge, the better summarization result it is considered to be. 
In addition, as a summary, to show the commonality and the rareness of a path among the segments, we annotate each edge with its appearance frequency in the segments.

\topic{Minimum \opsumgraph}: 
\opsum\ combines \eat{segment }vertices in their equivalence classes and ensures the paths in the summary \eat{output summary graph }satisfy above conditions. Next we define a valid summary graph.

Given \eat{a set of segments $\text{\segmentset}=\{\text{\segsubgraphFullWithI}\}$, and a \opsum$(\text{\propagg}, \text{\provtypehop{k}})$}a \opsum$(\text{\segmentset}, \text{\propagg}, \text{\provtypehop{k}})$ query, 
 a \ul{\emph{provenance summary graph}}, \opsumgraph$(\psgvertex, \psgedge, \psgvpropfunc, \psgepropfunc)$, is a directed acyclic graph, where 
\begin{enumerate}[leftmargin=18pt,label=\emph{\alph*})]{}
\item each $\mu \in \psgvertex$ represents a subset of an equivalence class $\mu \subseteq [v]$ w.r.t. \vertexeqrel\ over \segmentsetallv, 
and one segment vertex $v$ can only be in one \opsumgraph\ vertex $\mu$, i.e., $\forall \mu_m, \mu_n \in \psgvertex$, $\mu_m \cap \mu_n = \emptyset$; 
the vertex label function $\psgvpropfunc: \psgvertex \mapsto \text{\vertexeqrelpartition}$ maps a \opsumgraph\ vertex to its equivalence class; 
\item an edge $e_{m,n} = (\mu_m, \mu_n) \in \psgedge$ exists if there is a corresponding segment edge%
\eat{in some segment}, i.e., $\exists_{\text{\segsubgraph}_i}\ \mu_m \times \mu_n \cap \text{\sege}_i \neq \emptyset$; 
the edge label function $\psgepropfunc: \psgedge \mapsto [0,1]$ annotates the edge's frequencies over segments, i.e., $\psgepropfunc(e_{m,n}) = 
|\{\text{\segsubgraph}_i| \mu_m \times \mu_n \cap \text{\sege}_i \neq \emptyset\}|
/
|\text{\segmentset}|
$; 
\item there is a path \sumpath{m}{n} from $\mu_m$ to $\mu_n$ \emph{\textbf{iff}} $\exists_{\text{\segsubgraph}_i} v_s \in \mu_m \cap \text{\segv}_i \wedge v_t \in \mu_n \cap \text{\segv}_i$, there is a path \sumpath{s}{t} from $v_s$ to $v_t$ in $\text{\segsubgraph}_i$, and their path labels are the same $\text{\pathlabel}({\text{\sumpath{m}{n}}}) = \text{\pathlabel}({\text{\sumpath{s}{t}}})$. Note that in \opsumgraph, we use equivalence classes' canonical label (e.g., smallest vertex {\tt id}) as the vertex label in \pathlabel.
\end{enumerate} 

It is easy to see $\bigcup_i \text{\segsubgraph}_i$, \eat{the }union of all segments in \segmentset, is a valid \opsumgraph. \eat{However, w}We are interested in a concise summary with fewer vertices\eat{$|\psgvertex|$ the better}. 
The best \opsumgraph\ one can get is the optimal solution of the following problem.

\begin{problem}[Minimum \opsumgraph]
\label{problem_minpsg}
Given a set of segments \segmentset\ and a \opsum$($\segmentset$,$\propagg$,$\provtypehop{k}$)$ query, find the provenance summary graph \opsumgraph$(\psgvertex, \psgedge, \psgvpropfunc, \psgepropfunc)$ with minimum $|\psgvertex|$.
\end{problem}

\subsection{Query Evaluation}
Given \segmentset$=\{\text{\segsubgraphFullWithI}\}$, after applying \propagg\ and \provtypehop{k}, 
\opsumgraph\ $g_0 = \bigcup_i \text{\segsubgraph}_i$ is a labeled graph and contains all paths in segments; to find a smaller \opsumgraph, we have to merge vertices in \segmentsetallv$=\bigcup_i\text{\segv}_i$ in $g_0$ while keeping the \opsumgraph\ invariant, i.e., not introducing new paths. 
\eat{}
\eat{In order t}To describe merging conditions, we introduce \emph{trace equivalence} relations in a \opsumgraph. 
\emph{\textbf{a)}} in-trace equivalence (\intraceequal): If for every path $\text{\segpath{a}{u}}$ ending at $u \in \text{\segmentsetallv}$, there is a path $\text{\segpath{b}{v}}$ ending at $v \in \text{\segmentsetallv}$ with the same label, i.e., $\text{\pathlabel}(\text{\segpath{a}{u}}) = \text{\pathlabel}(\text{\segpath{b}{v}})$, we say $u$ is in-trace dominated by $v$, denoted as $u \text{\intracedominant} v$. The $u$ and $v$ are in-trace equivalent, written $u \text{\intraceequal} v$, iff $u \text{\intracedominant} v \wedge v \text{\intracedominant} u$. \emph{\textbf{b)}} out-trace equivalence (\outtraceequal): Similarly, if for every path starting at $u$, there is a path starting at $v$ with the same label, then we say $u$ is out-trace dominated by $v$, written $u \text{\outtracedominant} v$. $u$ and $v$ are out-trace equivalent, i.e., $u \text{\outtraceequal} v$, iff $u \text{\outtracedominant} v \wedge v \text{\outtracedominant} u$.

\begin{lemma}
\label{lm:merge_condition}
Merging $u$ to $v$ does not introduce new paths, if and only if \emph{1)} $u \text{\intraceequal} v$, or \emph{2)} $u \text{\outtraceequal} v$, or \emph{3)} $u \text{\intracedominant} v \wedge u \text{\outtracedominant} v$.
\end{lemma}

The lemma defines a partial order over the vertices in \segmentsetallv. By applying the above lemma, we can merge vertices in a \opsumgraph\ greedily until no such pair exist, then we derive\eat{the minimum} a minimal \opsumgraph. 
However, the problem of checking in-/out-trace equivalence is PSPACE-complete~\cite{trace_eq@stoc73}, which implies that the decision and optimization versions of the minimum \opsumgraph\ problem are PSPACE-complete.

\begin{theorem}
Minimum \opsumgraph\ is PSPACE-complete.
\end{theorem}

Instead of checking trace equivalence, we use \emph{simulation} relations as its approximation~\cite{comp_simulation@focs95,trace_approx@icdt99}, which is less strict than bisimulation and can be computed efficiently in O($|\text{\segmentsetallv}||\text{\segmentsetalle}|$) in \opsumgraph\ $g_0$~\cite{comp_simulation@focs95}. A vertex $u$ is in-simulate dominated by a vertex $v$, written $u \text{\insimulate} v$, if \emph{\textbf{a)}} their label \eat{in \opsumgraph\ }is the same, i.e., $\psgvpropfunc(u) = \psgvpropfunc(v)$ and \emph{\textbf{b)}} for each parent $p_u$ of $u$, there is a parent $p_v$ of $v$, s.t., $p_u \text{\insimulate} p_v$.
We say $u$, $v$ in-simulate each other, $u \text{\insimulationeq} v$, iff $u \text{\insimulate} v \wedge v \text{\insimulate} u$. Similarly, $u$ is out-simulate dominated (\outsimulate) by $v$, if $\psgvpropfunc(u) = \psgvpropfunc(v)$ and for each child $c_u$ of $u$, there is a child of $c_v$ of $v$, s.t., $c_u \text{\outsimulate} c_v$; and $u$, $v$ out-simulate each other $u \text{\outsimulationeq} v$ iff they out-simulate dominate each other. 
\eat{}
Note that a binary relation $r_a$ \emph{approximates} $r_b$, if $(e_i, e_j) \in r_a$ implies $(e_i, e_j) \in r_b$~\cite{trace_approx@icdt99}. In other words, if $(u,v)$ in-/out-simulates each other, then $(u,v)$ is in-/out-trace equivalence. By using simulation instead of trace equivalence in Lemma~\ref{lm:merge_condition} to merge\eat{ as the merge condition}, we can ensure the invariant. 

\begin{lemma}
\label{lm:approx_merge_condition}
If \emph{1)} $u \text{\insimulationeq} v$, or \emph{2)} $u \text{\outsimulationeq} v$, or \emph{3)} $u \text{\insimulate} v \wedge u \text{\outsimulate} v$, merging $u$ to $v$ does not introduce new paths.
\end{lemma}

We develop the \opsum\ algorithm by using the partial order derived from Lemma~\ref{lm:approx_merge_condition} merge condition in a \opsumgraph\ (initialized as $g_0$) to merge the vertices. To compute \insimulate\ and \outsimulate, we apply the similarity checking algorithm in~\cite{comp_simulation@focs95} twice in O($|\text{\segmentsetallv}||\text{\segmentsetalle}|$) time. 
\eat{}
From Lemma~\ref{lm:approx_merge_condition}, we can ensure there is no new path introduced, and the merging operation does not remove paths, so \opsum\ algorithm finds a valid \opsumgraph. Note\eat{ that} unlike Lemma~\ref{lm:merge_condition}, as the reverse of Lemma~\ref{lm:approx_merge_condition} does not hold, so we may not be able to find the minimum \opsumgraph, as there may be $(u, v)$ is in trace equivalence but not in simulation.

\section{Experimental Evaluation}
\label{sec:exp}

In this section, we study the proposed operators and techniques comprehensively. 
All experiments are conducted on a Ubuntu Linux 16.04 machine with an 8-core 3.0GHz AMD FX-380 processor and 16GB memory. For the backend property graph store, we use Neo4j 3.2.5 community edition in embedded mode and access it via its Java APIs. \eat{\provdb}Proposed query operators are implemented in Java in order to work with Neo4j APIs. To limit the performance impact from the Neo4j, we always use \eat{the node }{\tt id} to seek the nodes, which can be done in constant time in Neo4j's physical storage. Unless specifically mentioned, the page cache for Neo4j is set to 2GB and the JVM version is 1.8.0\_25 and -Xmx is set to 4GB. 

\topic{Dataset Description}: Unless lifecycle management systems\eat{ (e.g., Ground~\cite{ground@cidr17}, \provdb~\cite{provdb@hilda17})} are used by the practitioners for a long period of time, it is difficult to get real-world provenance graph from data science teams. \eat{Though using VCS (e.g., \cmd{git}) is common practice, VCS repositories only consist of versions of artifacts, but not the activities that occurred between commits. }Publicly available real-world \prov\ provenance graph datasets in various application domains~\cite{agg_lucmoreau@gam15} are very small (KBs). 
We instead develop several synthetic \prov\ graph generators to examine different aspects of the proposed operators. 
The datasets and the generators are available online\footnote{Datasets: {\scriptsize{\url{http://www.cs.umd.edu/~hui/code/provdbquery}}}}.

\subtopic{(a) Provenance Graphs \& \opseg\ Queries:}
To study the efficiency of \opseg, we generate a provenance graphs dataset (\syntheticpg) for collaborative analytics projects by mimicking a group of project members performing a sequence of activities\eat{ in a lifecycle management system}. 
Each project artifact has many versions and each version is an entity in the graph. 
An activity uses one or more input entities and produces one or more output entities.

\begin{figure*}[!t]
\centering{
\subfigure[Varying Graph Size $N$]{
  \includegraphics[totalheight=0.168\linewidth,trim=0 0 2 2]{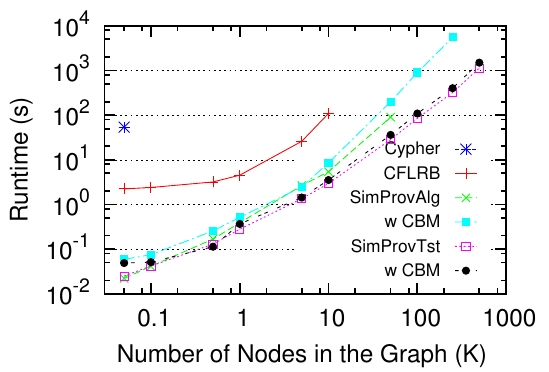} 
  \label{fig:exp1}
}
\subfigure[Varying Selection Skew $s_e$]{
  \includegraphics[totalheight=0.160\linewidth,trim=5 5 0 0]{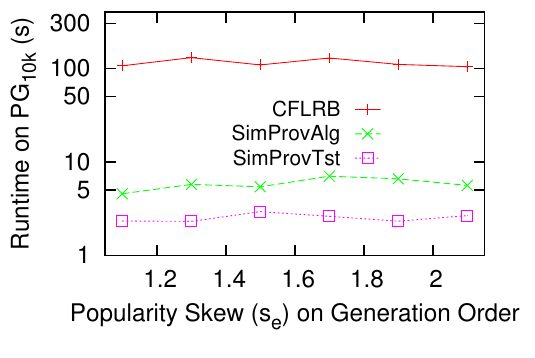} 
  \label{fig:exp2}
}
\subfigure[Varying Activity Input $\lambda_i$]{
  \includegraphics[totalheight=0.160\linewidth,trim=10 5 0 0]{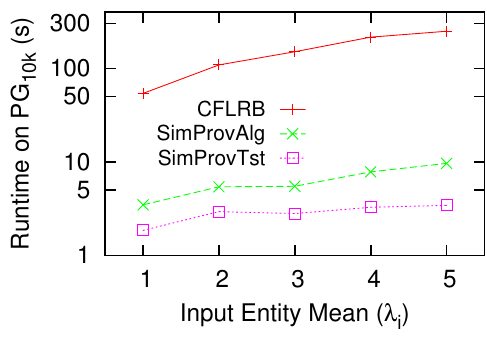}
  \label{fig:exp3}
}
\subfigure[Effectiveness of Early Stopping]{
  \includegraphics[totalheight=0.160\linewidth,trim=0 5 0 0]{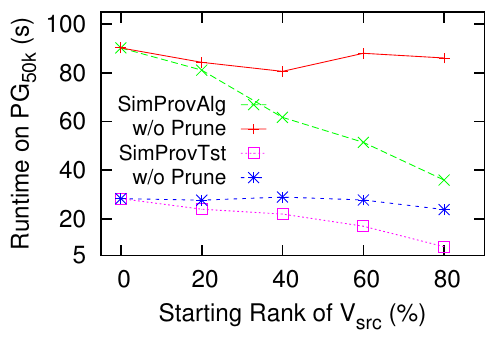}
  \label{fig:exp4}
}
\\\vspace{-3.5pt}
\subfigure[Varying Concentration $\alpha$]{
  \includegraphics[totalheight=0.165\linewidth,trim=0 0 5 4.5]{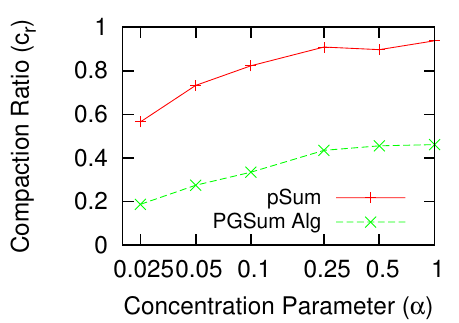} 
  \label{fig:sum_exp1}
}
\subfigure[Varying Activity Types $k$]{
  \includegraphics[totalheight=0.165\linewidth,trim=3 0 5 4.5]{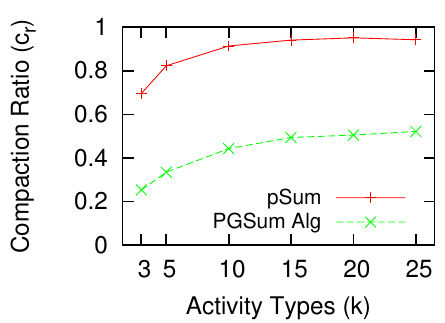} 
  \label{fig:sum_exp2}
}
\subfigure[Varying Number of Activities $n$]{
  \includegraphics[totalheight=0.165\linewidth,trim=3 0 5 4.5]{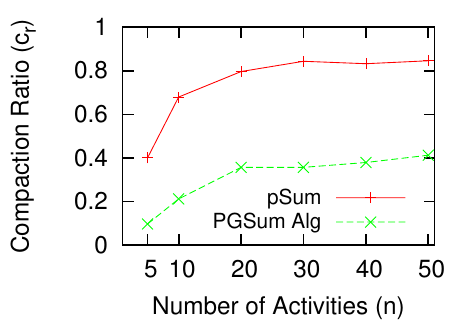}
  \label{fig:sum_exp3}
}
\subfigure[Varying $|\text{\segmentset}|$]{
  \includegraphics[totalheight=0.165\linewidth,trim=3 0 5 4.5]{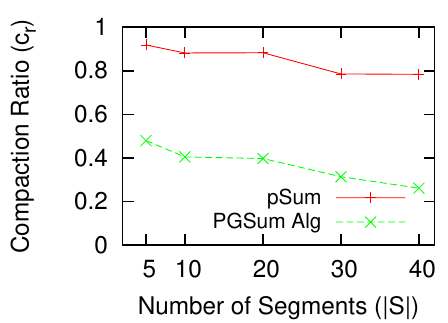}
  \label{fig:sum_exp4}
}
}
\vspace{-11pt}
\caption{Efficiency and Effectiveness Evaluation Results for \opseg\ and \opsum\ Algorithms}
\label{fig:optseg_eff}
\vspace{-17.5pt}
\end{figure*}

To elaborate, given $N$, the number of vertices in the output graph, we introduce
$|\text{\agent}| = \floor*{\log(N)}$ agents. To determine who performs the next activity, we use a Zipf distribution with skew $s_w$ to model their work rate.
Each activity is associated with an agent and uses $1+m$ input entities and generates $1+n$ output entities. $m$ and $n$ are generated from two Poisson distributions with mean $\lambda_i$ and $\lambda_o$ to model different input and output size. 
In total, the generator produces 
$|\text{\activity}| = \floor*{{N}/{(2+\lambda_o)}}$ 
activities, so that at the end of generation, the sum of entities $|\text{\entity}|$, activities $|\text{\activity}|$ and agents $|\text{\agent}|$ is close to $N$. 
The $m$ input entities are picked from existing entities; the probability of an entity being selected is modeled as the pmf of a Zipf distribution with skew $s_e$ at its rank in the reverse order of being. If $s_e$ is large, then the activity tends to pick the latest generated entity, while $s_e$ is small, an earlier entity has better chance to be selected. 
\eat{The $n$ output entities are always new entities, which would be the first version of an artifact, or a new version of an existing artifact. For the latter, we add a derivation edge to an ancestor entity uniformly. }

We use the following values as default for the parameters: $s_w=1.2$, $\lambda_i = 2$, $\lambda_o = 2$, and $s_e = 1.5$. We refer $\text{\syntheticpg}_{n}$ as the graph with about $n$ vertices.
In \syntheticpg\ graphs, 
we pick pairs 
(\segSource, \segDestination) as \opseg\ queries to evaluate. Unless specifically mentioned, given a \syntheticpg\ dataset, \segSource\ are the first two entities, and \segDestination\ are the last two entities, as they are always connected by some path and the query is the most challenging \opseg\ instance. In one evaluation, we vary \segSource\ to show the effectiveness of the proposed pruning strategy.

\subtopic{(b) Similar Segments \& \opsum\ Queries:}
To study the effectiveness of \opsum, we design a synthetic generator (\syntheticsg) with the ability to vary shapes of conceptually similar provenance graph segments. In brief, the intuition is that as at different stages of the project, the stability of the underlying pipelines tends to differ, the effectiveness of summary operator could be affected; e.g., at the beginning of a project, many activities (e.g., clean, plot, train\eat{ data}) would happen after another one in no particular order, while at later stages, there are more likely to be stable pipelines, i.e., an activity type (e.g., preprocessing) is always followed by another activity type (e.g., train). For \opsum, the former case is more challenging\eat{ than the latter one}.

In detail, we model a segment as a Markov chain with $k$ states and a transition matrix $M \in [0,1]^{k\times k}$ among states. Each row of the transition matrix is generated from a Dirichlet prior with the concentration parameter $\vec{\alpha}$, i.e., the $i$th row is a categorical distribution for state $i$; each $M_{ij}$ represents the probability of moving to state $j$, i.e., pick an activity of type $j$. We set a single $\alpha$ for the vector $\vec{\alpha}$; for higher $\alpha$, the transition tends to be a uniform distribution, while for lower $\alpha$, the probability is more concentrated, i.e., fewer types of activities would be picked from.
\eat{}
Given a transition matrix, we can generate a set of segments \segmentset, each of which consists of $n$ activities labeled with $k$ types, derived step by step using the transition matrix. For the input/output entities\eat{ and edges} of each activity, we use $\lambda_i$, $\lambda_o$, and $s_e$ the same way in \syntheticpg, and all introduced entities have the same equivalent class label.

\eat{
We vary $\alpha$, $|\text{\segmentset}|$, $k$ and $n$ to study the \opsum\ effectiveness on different sets of segments. A \opsum\ query is applied on each \segmentset, and produces a \opsumgraph. The effects of property aggregation and provenance types are reflected in the above label assignment process.
}

\topic{Segmentation Operator:}
We compare our algorithms \opsegAlg\ and \opsegAlgTst\ with the state-of-the-art general CFLR\eat{context-free language reachability} algorithm, \baselineAlg~\cite{chaudhuri_subcublic@popl08}, and the Cypher query in Sec.~\ref{subsec:seg_op} in Neo4j. 
\eat{It uses bit-based set operations to improve the Reps'~\cite{reps_cflr_baseline@tcs00} dynamic programming algorithm. }
We implement the fast set using \emph{\textbf{a)}} Java BitSet \eat{in order }to have constant random access time, \emph{\textbf{b)}} RoaringBitMap which is the state-of-the-art compressed bitmap ({\sc Cbm}) with slower random access but better memory usages~\cite{roaringmap1,roaringmap2}. 
\eat{We also compare with the Cypher query in Sec.~\ref{subsec:seg_op} in Neo4j. }

\subtopic{(a) Varying Graph Size $N$}: 
In Fig.~\ref{fig:exp1}, we study the scalability of all algorithms. $x$ axis denotes $N$ of the \syntheticpg\ graph, while $y$ axis shows the runtime in seconds to finish the \opseg\ query. Note the figure is log-scale for both axes. As we see, \opsegAlg\ and \opsegAlgTst\ run at least one order of magnitude faster than \baselineAlg\ on all \syntheticpg\ datasets, due to the utilization of the properties of the grammar and efficient pruning strategies. 
Note \baselineAlg\ runs out of memory on $\text{\syntheticpg}_{50k}$ due to much faster growth of the worklist\eat{ than \opsegAlg}, as the normal forms introduce an extra level; \opsegAlg\ without {\sc Cbm} runs out of memory on $\text{\syntheticpg}_{100k}$ due to O($n^2/\log(n)$) space complexity and 32bit integer interface in BitSet. 
\eat{}
With {\sc Cbm}, \eat{except \baselineAlg, }\baselineAlg\ still runs out of memory on $\text{\syntheticpg}_{50k}$ due to the worklist growth, Both \similarPathPatternRule\ algorithms reduce memory usages however become several times slower;  In particular, \opsegAlg\ uses 64bit RoaringBitMap\eat{space usage drops to O($n/\log(n)$)} and is scalable to larger graphs. For very large graphs, disk-based batch processing is needed and Datalog evaluation can be used (Sec.~\ref{sec:related_work}).

\opsegAlg\ runs slightly faster than \opsegAlgTst\ for small instances while becomes much slower for large instances, e.g., $\text{\syntheticpg}_{50k}$, it is 3x slower than \opsegAlgTst\ for the query. The reason \eat{that small instances \opsegAlg\ slightly faster }is the because \opsegAlgTst\ run $|\text{\segDestination}|$ times on the graph and each run's performance gain is not large enough. When the size of the graph instance increases, the superiority of the \opsegAlgTst\ by using the transitivity property becomes significant. 

On the other hand, the Cypher query can only return correct result for the very small graph $\text{\syntheticpg}_{50}$ and takes orders of magnitude longer. Surprisingly, even for small graph  $\text{\syntheticpg}_{100}$, it runs over 12 hours and we have to terminate it. By using its query profile tool, we know that Neo4j uses a path variable to hold all paths and joins them later which is exponential w.r.t. the path length and average out-degree. Due to the expressiveness of the path query language, the grammar properties cannot be used by the query planer. 

\subtopic{(b) Varying Input Selection Skew $s_e$}: 
Next, in Fig.~\ref{fig:exp2}, we study the effect of different \eat{input entities }selection behaviors on $\text{\syntheticpg}_{10k}$. The $x$ axis is $s_e$ and the $y$ axis is the runtime in seconds in log-scale. In practice, 
\eat{some analytics activities tend to try many model alternatives to get the best performance for an analytics task, e.g., through a grid search over hyperparameters, or changing a neural network architecture; while there are other analytics activities having\eat{where there are} long chains of data transformation pipelines, e.g., feature engineering efforts. The former}some types of projects tend to always take an early entity as input (e.g., dataset, label), while \eat{the latter}some others tend to take new entities (i.e., the output of the previous \eat{pipeline }step) as inputs. Tuning $s_e$ in opposite directions can mimic those project behaviors\eat{, as it tunes the probability of earlier entities been selected as inputs}. In Fig.~\ref{fig:exp2}, we vary $s_e$ from $1.1$ to $2.1$, and the result is quite stable for \opsegAlg, \opsegAlgTst\ and \baselineAlg, which implies the \eat{query formulation and techniques}algorithms can be applied to different project types with similar performance. 

\subtopic{(c) Varying Activity Input Mean $\lambda_i$}: 
We study the effect of varying density of the graph in Fig.~\ref{fig:exp3} on $\text{\syntheticpg}_{10k}$. The $x$ axis varies the mean $\lambda_i$ of the number of input entities. The $y$ axis shows the runtime in seconds. Having a larger $\lambda_i$, the number of $|\text{\used}|$ edges will increase linearly, thus the\eat{ algorithms} runtime increases\eat{linearly} as well. In Fig.~\ref{fig:exp3}, we see \opsegAlg\ grows much more slowly than \baselineAlg. Due to the pruning strategies, the growth in worklist is reduced.\eat{ utilization is avoided in \opsegAlg.} \opsegAlgTst\ has the best performance\eat{further improves the \opsegAlg\ }due to the utilization of the transitivity.

\subtopic{(d) Effectiveness of Early Stopping}: 
The above evaluations all use the most challenging \opseg\ query on start and end entities. In practice, we expect the users will ask queries whose result they can understand by simple visualization. \baselineAlg\ and general CFL don't have early stopping properties. \eat{\opsegAlg\ and \opsegAlgTst\ }Our algorithms use the temporal constraints of the provenance graph to support early stopping growing the result. In Fig.~\ref{fig:exp4}, we vary the \segSource\ \eat{of a \opseg\ query }and study the performance on $\text{\syntheticpg}_{50k}$. The $x$ axis is the starting position among all the entities, e.g., $x=20$ means \segSource\ is selected at the end of 20\% percentile w.r.t. the ranking of the order of being. \eat{The $y$ axis is the runtime in seconds. }As we can see, the shorter of the temporal gap between \segSource\ and \segDestination, the shorter our algorithms' \eat{the \opsegAlg\ and \opsegAlgTst\ }runtime. \eat{By utilizing}Using the property of \prov\ graphs, we get better performance empirically even though the worst case complexity does not change. 

\topic{Summarization Operator:}
Given a \segmentset$=\{\text{\segsubgraphFullWithI}\}$, \opsum\ generates a precise summary graph \opsumgraph$(\psgvertex, \psgedge)$ by definition. Here we study its effectiveness in terms of conciseness. We use the compaction ratio defined as $c_r = |\psgvertex|/|\bigcup_i\text{\segv}_i|$. As there are few graph query result summarization techniques available, the closest applicable algorithm we can find is\eat{ in our study, we compare with} \sumbaseline~\cite{yinghui_summary@pvldb13} which is designed for summarizing a set of graphs from keyword search graph queries. \sumbaseline\ works on undirected graphs and preserves path among keyword pairs and was shown to be more effective than summarization techniques on one large graph, e.g., SNAP~\cite{yytian_summary@sigmod08}. To make \sumbaseline\ work on \opseg\ segments, we introduce a conceptual (\emph{start}, \emph{end}) vertex pair as the keyword vertices, and let the \emph{start} vertex connect to all vertices in \segmentset\ having $0$ in-degree, and similarly let the \emph{end} vertex connect to all vertices having $0$ out-degree. In the rest of the experiments, by default, $\alpha=0.1$, $k=5$, $n=20$ and $|\text{\segmentset}| = 10$, and y-axis denotes\eat{the compaction ratio} $c_r$\eat{ in all figures}.

\subtopic{(a) Varying Transition Concentration $\alpha$}: 
In Fig.~\ref{fig:sum_exp1}, we change the concentration parameter to mimic segment sets at various stage of a project with different stableness. $x$ axis denotes the value of $\alpha$ in log-scale. Increasing $\alpha$, the transition probability tends to be uniform, in other words, the pipeline is less stable, and paths are more likely be different, so the vertex pairs which would be merged become infrequent. 
As we can see, \opsum\ algorithm always performs  better than \sumbaseline, and the generated \opsumgraph\ is about half the result produced by \sumbaseline, as \sumbaseline\ cannot combine some \intraceequal\ and \outtraceequal\ pairs, which are important for workflow graphs.\eat{ The same finding is consistent in other experiments.}

\subtopic{(b) Varying Activity Types $k$}: 
Next, in Fig.~\ref{fig:sum_exp2}, we vary the possible transition states, which reflects the complexity of the underlying pipeline. It can also be viewed as the effect of using property aggregations on activities (e.g., distinguish the commands with the same name but different options). Increasing $k$ leads to more different path labels, as shown in the Fig.~\ref{fig:sum_exp2}, and it makes the summarization less effective. Note that when varying $k$, the number of activities $n$ in a segment is set to be $20$, so the effect of $k$ on compaction ratio tends to disappear when $k$ increases.

\subtopic{(c) Varying Segment Size $n$}: 
We vary the size of each segment $n$ when fixing $\alpha$ and $k$ to study the performance of \opsum. Intuitively, the larger the segment is, the more intermediate vertices there are. The intermediate vertices are less likely to satisfy the merging conditions due to path constraints. In Fig.~\ref{fig:sum_exp3}, $c_r$\eat{the compaction ratio} increases as the input instances are more difficult. 

\subtopic{(d) Varying Number of Segments $|\text{\segmentset}|$}: 
With all the shape parameters set ($\alpha=0.25$), we increase the number of similar segments. As the segments are derived by\eat{generated from} the same transition matrix, they tend to have similar paths. \eat{As shown i}In Fig.~\ref{fig:sum_exp4}, $c_r$\eat{the compaction ratio} becomes better when more segments are given as input.

\section{Related Work}
\label{sec:related_work}

\topic{Provenance Systems:}
Provenance studies\eat{ in the literature} can be roughly categorized in two types: \eat{often distinguish themselves by }data provenance\eat{ (a.k.a. fine-granularity)} and workflow provenance\eat{
    (a.k.a. coarse-granularity)}. \eat{On one hand, d}Data provenance is discussed in dataflow \eat{data-centric systems having dataflow query facilities}systems, such as RDBMS, Pig Latin, and
    Spark~\cite{survey_chiew@ftdb09,lipstick@pvldb11,titian@pvldb15}\eat{. On the other hand, }, while workflow provenance studies address complex interactions among high-level conceptual components in
    various computational tasks, such as scientific workflows, business processes, and \eat{system}cybersecurity~\cite{workflow_survey@cse08,bpql_milo@vldb06,linuxprov_abates@atc15}. 
Unlike \eat{retrospective }query facilities in scientific workflow provenance systems~\cite{workflow_survey@cse08}, their processes are predefined in \emph{workflow skeletons}, and multiple executions
generate different instance-level provenance \emph{run graphs} and have clear boundaries. Taking advantages of the skeleton\eat{definition}, there are lines of research \eat{to aid}for advanced ancestry
query processing\eat{are important topics of study}, such as defining user views over such skeleton to aid queries on verbose run graphs~\cite{zoom_penn@icde08}, querying reachability\eat{executing reachability query} on the run
graphs efficiently~\cite{reachability_optimallabeling_upenn@sigmod10}, storing run graphs generated by the skeletons compactly~\cite{compression_bertram@edbt09}, and using visualization as examples to
ease query construction~\cite{visualization@vis05}. 

Most relevant work is querying evolving script provenance~\cite{ingestions_noworkflow@ipaw14,noworkflow_evolution@ipaw16}. Because script
executions form clear \eat{run graph }boundary, query facilities to visualize and difference execution run graphs are proposed. In our context,
           as there are no clear boundaries of run graphs, it is crucial to design query facilities allowing the user to express the logical run graph segments and specify the boundary conditions first. Our method can also be applied on script provenance by segmenting within and summarizing across evolving run graphs.

\topic{Data Science Lifecycle Management:}
Recently, there is emerging interest in developing systems for managing different aspects in the modeling lifecycle, such as building modeling lifecycle platforms~\cite{google_tfx@kdd17}, accelerating iterative modeling process~\cite{columbus_tods16p}, managing developed models~\cite{modeldb_HILDA16p,modelhub@icde17}, organizing lifecycle provenance and metadata~\cite{ground@cidr17,provdb@hilda17,amazon_metadata@learningsys17},  auto-selecting models~\cite{model_selection_arun_record15p}, hosting pipelines and discovering reference models~\cite{openml@kddexp13,discovery@deem17}, and assisting collaboration~\cite{kandogan2015labbookp}. 
Issues of querying evolving and verbose provenance effectively are not considered in that work.

\topic{Context Free Language \& Graph Query:}
Parsing CFL on graphs and using it as query primitives has been studied in early theory work~\cite{valiant_subcublic@jcss75,cflr_to_datalog@pods90}, later used widely in programming analysis~\cite{reps_pa_reachability@islp97} and other domains such as bioinformatics~\cite{alon@amia02} which requires high expressiveness language to constrain paths. Recently it is discussed as a graph query language~\cite{cflpq@icdt14} and SPARQL extension~\cite{cflsparql@iswc16} in graph databases. In particular, CFLR is a general formulation of many program analysis tasks on graph representations of programs. Most of the CFL used in program analysis is a Dyck language for matching parentheses~\cite{reps_pa_reachability@islp97}. On provenance graphs, our work is the first to use CFL to constrain the path patterns to the best of our knowledge. CFL allows us to capture path similarities and constrain lineages in evolving provenance graphs. We envision many interesting provenance queries would be expressed in CFLR and support human-in-the-loop introspection of the underlying workflow.

Answering CFLR on graphs in databases has been studied in~\cite{cflr_to_datalog@pods90}, and shown equivalent to evaluating Datalog chain programs. Reps~\cite{reps_pa_reachability@islp97,reps_cflr_baseline@tcs00} describe a cubic time algorithm to answer CFLR and is widely used in program analysis. Later it is improved in~\cite{chaudhuri_subcublic@popl08} to subcubic time. Because \eat{Though }general CFLR is generalization of CFL parsing, it\eat{ which} is difficult to improve~\cite{reps_pa_reachability@islp97}. \eat{Due to its importance in PL, faster algorithms for\eat{ the} Dyck language reachability on specific data models and tasks are discussed in the programing language community}
On specific data models and tasks, faster algorithms for Dyck language reachability are discussed in the PL community~\cite{treecfl@esop09,fastcfl@pldi13}. Our work can be viewed as utilizing provenance graph properties and rewriting CFG to improve CFLR evaluation. 

\topic{Query Results \& Graph Summarization:}
Most work on graph summarization~\cite{summary_tutorial@pvldb17} focuses on finding smaller representations for a very large graph by methods such as compression~\cite{rastogi_mdl@sigmod08}, attribute-aggregation~\cite{yytian_summary@sigmod08} and bisimulation~\cite{bisimulation@icde02}; while there are a few works aiming at combining a set of query-returned trees~\cite{xmlresults@sigmod08} or graphs~\cite{yinghui_summary@pvldb13} to form a compact representation. 
Our work falls into the latter category\eat{, and is tailored for \opseg\ segments which consist of similar or alternative steps among a set of entities of interest}. Unlike other summarization techniques, 
our operator is designed for provenance graphs which include multiple types of vertices rather than a single vertex type~\cite{graph_cube@sigmod11}; it works on query results rather than entire graph structure~\cite{yytian_summary@sigmod08,rastogi_mdl@sigmod08,agg_lucmoreau@gam15}; the summarization requirements are specific to provenance graphs rather than returned trees~\cite{xmlresults@sigmod08} or keyword search results~\cite{yinghui_summary@pvldb13}. We also consider property aggregations and provenance types \eat{in our query constructs }to allow tuning provenance meanings, which is not studied before to the best of our knowledge.

\section{Conclusion}
\label{sec:conclusion}

We described the key challenges in querying provenance graphs generated in evolving workflows without predefined skeletons and clear boundaries, such as the ones collected by lifecyle management systems in collaborative analytics projects. 
At query time, as the users only have partial knowledge about the ingested provenance, due to the schema-later nature of the
    properties, multiple versions of the same files, unfamiliar artifacts introduced by team members, and enormous provenance records
        collected continuously. Just using standard graph query model is highly ineffective in utilizing the valuable information. We presented two \eat{high-level }graph query operators to address the verboseness and evolving nature of such provenance graphs. First, \eat{we introduced a graph}the segmentation operator \eat{that }allows the users to only provide the vertices they are familiar with and then induces a subgraph representing the retrospective provenance of the vertices of interest. 
        We formulated the semantics of such a query in a context free language, and developed efficient algorithms on top of a property graph backend. Second, \eat{we described a graph}the summarization operator\eat{ that} combines the results of multiple segmentation queries and preserves provenance meanings to help users understand similar and abnormal behavior\eat{ in those conceptually similar segments and allows to tune the provenance meanings}. \eat{with multi-resolution capabilities. }Extensive experiments on synthetic provenance graphs with different project characteristics show the operators and evaluation techniques are effective and efficient. The operators are also applicable for querying provenance graphs generated in other scenarios where there are no workflow skeletons, e.g., cybersecurity and system diagnosis.

\newpage
\bibliographystyle{IEEEtran}
\bibliography{main}

\newpage
\appendix
\subsection{Notation Table} 
\label{apdx:notation}
We summarize the notations used for defining operators' semantics in the paper in Table~\ref{tb:notations}.
\begin{table}[h!]
\centering
{\footnotesize{
\begin{tabular}{l | l | l | l}
\toprule
\multicolumn{4}{c}{Provenance Graph: $\text{\provgraph}(\text{\vertexset}, \text{\edgeset}, \text{\vlabelfunc}, \text{\elabelfunc}, \text{\vpropfunc},\text{\epropfunc})$} \\
\hline
\provgraph & \prov\ graph & \entity & Entity \\
\vertexset\eat{$=$\entity$\cup$\activity$\cup$\agent} & Vertices (\entity$\cup$\activity$\cup$\agent) & \activity & Activity \\
\edgeset\eat{$=$\used$\cup$\wasGeneratedBy$\cup$\wasAssociatedWith$\cup$\wasAttributedTo$\cup$\wasDerivedFrom} & Edges (\used$\cup$\wasGeneratedBy$\cup$\wasAssociatedWith$\cup$\wasAttributedTo$\cup$\wasDerivedFrom) & \agent & Agent \\
\vlabelfunc\eat{ $:$ \vertexset$\mapsto\{$\entity$,$\activity$,$\agent$\}$} & Vertex label & \used & used \\
\elabelfunc\eat{ $:$ \edgeset$\mapsto\{$\used$,$\wasGeneratedBy$,$\wasAssociatedWith$,$\wasAttributedTo$,$\wasDerivedFrom$\}$} & Edge label & \wasGeneratedBy & wasGeneratedBy \\
\property &  Property types & \wasAssociatedWith & wasAssociatedWith \\
\vpropfunc\eat{$:$ \vertexset $\times $ \property $\mapsto \mathcal V$} & Vertex Property & \wasAttributedTo & wasAttributedTo \\
\epropfunc\eat{$:$ \edgeset $\times $ \property $\mapsto \mathcal V$} & Edge Property & \wasDerivedFrom & wasDerivedFrom \\
\midrule
\multicolumn{4}{c}{Segmentation Operator: \opseg$(\text{\segSource},\text{\segDestination},\text{\boundaryCriteria})$} \\
\hline
\segSource\eat{ $\subseteq$ \entity} &  Source entities & \segsubgraphFull & Segment graph \\
\segDestination\eat{ $\subseteq$ \entity} & Destination entities & \segpath{$v_i$}{$v_j$} & Path \\
\segInduced & Induced vertices & \segsubpath{$v_i$}{$v_j$} & Path segment \\
\boundaryCriteria & Boundary criteria & \inv{\segpath{a}{b}},\inv{\segsubpath{a}{b}} & Inverse path (segment) \\
\pathlabel & Path label function & \similarPathPatternRule & Context free grammar \\
$\varepsilon$ & Empty word & \cfglanguage{\similarPathPatternRule} & Context free language \\
\midrule
\multicolumn{4}{c}{Summarization Operator: \opsum$(\text{\segmentset}, \text{\propagg}, \text{\provtypehop{k}})$} \\
\hline
\propagg & Property Aggregation & \segmentset & A set of segments  \\
\provtypehop{k} & Provenance Type & $\psgvertex$ & \opsumgraph\ vertices \\
\vertexeqrel & Vertex Eqv. class & $\psgvpropfunc$ & \opsumgraph\ vertex label \\
\opsumgraph & Summary graph & $\psgepropfunc$ & \opsumgraph\ edge label \\ 
\bottomrule
\end{tabular}
}}
\caption{Summary of Notations}
\label{tb:notations}
\vspace{-25pt}
\end{table}

\begin{figure}[b!]
\begin{equation*}
{\small{
\begin{aligned}
  r_0: \text{\sc Qd} &\rightarrow\quad v_j                                        & \forall v_j \in &\text{\segDestination}                                      & \\
  r_1: \text{\sc Lg} &\rightarrow\quad \text{\wasGeneratedByInv}\ \text{\sc Qd}   & r_3: \text{\sc La} &\rightarrow\quad \text{\activity}\ \text{\sc Rg}  & r_6: \text{\sc Ru} &\rightarrow\quad \text{\sc Lu}\ \text{\used}    \\
   &\quad|\quad  \text{\wasGeneratedByInv}\ \text{\sc Re}       & r_4: \text{\sc Ra} &\rightarrow\quad \text{\sc La}\ \text{\activity}  & r_7: \text{\sc Le} &\rightarrow\quad \text{\entity}\ \text{\sc Ru}  \\
  r_2: \text{\sc Rg} &\rightarrow\quad \text{\sc Lg}\ \text{\wasGeneratedBy}      & r_5: \text{\sc Lu} &\rightarrow\quad \text{\usedInv}\ \text{\sc Ra}   & r_8: \text{\sc Re} &\rightarrow\quad \text{\sc Le}\ \text{\entity}  \\
\end{aligned}
}}
\end{equation*}
\vspace{-15pt}
\caption{\similarPathPatternRule\ Normal Form\eat{, \similarPathPatternRule\ $\rightarrow \text{\sc Re}$}. %
$\text{\sc Lg} \subseteq \text{\activity}\times\text{\entity}$; %
$\text{\sc Rg, La, Ra} \subseteq \text{\activity}\times\text{\activity}$; %
$\text{\sc Lu} \subseteq \text{\entity}\times\text{\activity}$; %
$\text{\sc Ru, Le, Re, Qd} \subseteq \text{\entity}\times\text{\entity}$. %
}
\label{fig:simprov_normalized}
\end{figure}

\subsection{Segmentation Operation}
\label{apdx:seg_algs}

\topic{\baselineAlg\ Algorithm for \cfglanguage{\similarPathPatternRule}-reachability}: 
\baselineAlg~\cite{chaudhuri_subcublic@popl08} (shown in Alg.~\ref{alg:cflrb}) is a subcubic algorithm to solve general CFLR problem. 
Given a CFG, \baselineAlg\ works on its normal form~\cite{intr_automata@3ed}, where each production has at most two RHS symbols, i.e., $N \rightarrow AB$ or $N \rightarrow A$. We show the normal form of \similarPathPatternRule\ in Fig.~\ref{fig:simprov_normalized} (domain of LHS of each production rule is shown in the caption). At a high level, the algorithm traverses the graph and uses grammar as a guide to find new production facts $N(i,j)$, where $N$ is a LHS nonterminal, $i,j$ are graph vertices, and the found fact $N(i,j)$ denotes that there is a path from $i$ to $j$ whose path label satisfies $N$. To elaborate, similar to BFS, it uses a worklist $W$ (queue) to track newly found fact $N(i,j)$ and a \emph{fast set} data structure $H$ with time complexity $O(n/log(n))$ for set diff/union and $O(1)$ for insert to memorize found facts. 

In the beginning, all facts $F(i,j)$ from all single RHS symbol rules $F \rightarrow A$ are enqueued. In \similarPathPatternRule\ case ($r_0$: {\sc Qd} in Fig.~\ref{fig:simprov_normalized}), each $v_j \in$\segDestination\ is added to $W$ as $\text{\sc Qd}(v_j, v_j)$. From $W$, it\eat{the algorithm} processes one fact $F(i,j)$ at a time until $W$ is empty. When processing a dequeued fact $F(i,j)$, if $F$ appears in any rule in the following cases:
\begin{equation*}
\begin{aligned}
& N(i,j)\rightarrow F(i,j); \\
& N(i,v)\rightarrow F(i,j)A(j,v); \\
& N(u,j)\rightarrow A(u,i)F(i,j)
\end{aligned}
\end{equation*}
the new LHS fact $N(i,v)$ is derived by set diff $\{v \in A(j,v)\}\setminus \{v \in N(i,v)\}$ or $N(u,j)$ by $\{u \in A(u,i)\}\setminus \{u \in N(u,j)\}$ in $H$. As in \similarPathPatternRule, only the later two cases are present, in Alg.~\ref{alg:cflrb} line 4~6 and line 7~9 show the details of the algorithm. Row and Col accesses outgoing and incoming neighbors w.r.t. to a LHS symbol and is implemented using the fast set data structure. Then the new facts of $N$ are added to $H$ to avoid repetition and $W$ to explore it later. Once $W$ is empty, the start symbol $L$ facts $L(i,j)$ in $H$ include all vertices pairs $(i,j)$ which have a path with label that satisfies $L$. If a grammar has $k$ rules, then the worst case time complexity is $O(k^3 n^3 /log(n)$ and $W$ ad $H$ takes $O(k n^2)$ space. If path is needed, a parent table would be used similar to BFS using standard techniques. In \similarPathPatternRule\ (Fig.~\ref{fig:simprov_normalized}), the start symbol is {\sc Re}, $\forall v_i \in \text{\segSource}$, $\text{\sc Re}(v_i, v_t)$ facts include all $v_t$, s.t. between them there is $\text{\pathlabel}({\text{\segsubpath{i}{t}}}) \in \text{\cfglanguage{\similarPathPatternRule}}$. 

\begin{algorithm}[t!] 
\caption{\baselineAlg: Subcubic time CFLR~\cite{chaudhuri_subcublic@popl08} for \similarPathPatternRule}
\label{alg:cflrb}
\begin{algorithmic}[1]
{\small{
\STATE $W \leftarrow H \leftarrow \{(v, \text{\sc Qd}, v) | v \in \text{\segDestination}\} $ 
\WHILE{$ W \neq \emptyset$}
  \STATE $(u, B, v) \leftarrow W.pop $
  \FOR{each production rule $A \rightarrow CB$} 
    \STATE $W \leftarrow H \leftarrow \{ (u', A, v) | u' \in \text{Col}(u, C) \setminus \text{Col}(v, A)\}$
  \ENDFOR
  \FOR{each production rule $A \rightarrow BC$}
    \STATE $W \leftarrow H \leftarrow \{ (u, A, v') | v' \in \text{Row}(v, C) \setminus \text{Row}(u, A)\}$
  \ENDFOR
\ENDWHILE
}}
\end{algorithmic}
\end{algorithm}

\topic{Proof of Lemma 1}: The proof of Lemma 1 is the following:
On \similarPathPatternRule\ normal form (Fig.~\ref{fig:simprov_normalized}), for $i \in [1,8]$, \baselineAlg\ derives $r_i$ LHS facts by a $r_{i-1}$ LHS fact dequeued from $W$ (Note it also derives $r_1$ from $r_8$). For $i \in \{1,2\}$, $r_i(u,v)$ uses \wasGeneratedBy\ edges in the graph during the derivation, e.g., from $r_8$ LHS $\text{\sc Re}$ to $r_1: \text{\sc Lg}(u,v) \rightarrow \text{\wasGeneratedByInv}(u, k)\ \text{\sc Re}(k, v)$. As $\text{\sc Re}(k,v)$ can only be in the worklist $W$ once, we can see that each 3-tuple $(u,k,v)$ is formed only once on the RHS and there are at most $|\text{\wasGeneratedBy}||\text{\entity}|$ of such 3-tuples. To make sure $\text{\sc Lg}(u,v)$ is not found before, $H$ is checked. If not using fast set but a $O(1)$ time procedure for each instance $(u,k,v)$, then it takes $O(|\text{\wasGeneratedBy}||\text{\entity}|)$ to produce the LHS; on the other hand, if using a \emph{fast set} on $u's$ domain \activity\ for each $u$, for each $\text{\sc Re}(k,v)$, $O({|\text{\activity}|}/{\log|\text{\activity}|})$ time is required, thus it takes $O({|\text{\activity}||\text{\entity}|^2}/{\log|\text{\activity}|})$ in total. Applying similar analysis on $r_5$ and $r_6$ using \used\ to derive new facts, we can see it takes $O({|\text{\entity}||\text{\activity}|^2}/{\log|\text{\entity}|})$ with fast set and $O(|\text{\used}||\text{\activity}|)$ without fast set. Finally $r_3, r_4$ and $r_7, r_8$ can be viewed as following a vertex self-loop edge and do not affect the complexity result. 

\subsection{Query Evaluation Discussion}
\topic{Validness of Segments} Validness of provenance graph is an important constraint~\cite{prov_constraints@w3c_tr13,prov_segmentation@tapp16}. In our system, the \opseg\ operator does not introduce new vertices or edge. As long as the original provenance graph is valid, the induced subgraph is valid.
However, at query time, the boundaries criteria could possibly let the operator result exclude important vertices. As an interactive system, we leave it to the user to adjust the vertex set of interest and boundary criteria in their queries.

\topic{Two-step approach Revisit}: For other purposes where the two-step approaches are not ideal, the exclusion constraints $\text{\boundaryCriteria}_v$ and $\text{\boundaryCriteria}_e$, and expansion criteria $\text{\boundaryCriteria}_x$ can be evaluated together using \baselineAlg, \opsegAlg\ and \opsegAlgTst\ with small modifications on the grammar. In \baselineAlg\, the label function $\mathcal{F}_v$ of $\text{\boundaryCriteria}_v$ can be applied at $r_0, r_3, r_4, r_7, r_8$ on \activity\ or \entity, while $\mathcal{F}_e$ of $\text{\boundaryCriteria}_e$ can be applied at rest of the rules involving \used\ and \wasGeneratedBy. For \opsegAlg\ and \opsegAlgTst, $\mathcal{F}_v$ and $\mathcal{F}_e$ can be applied together at $r'_1, r'_2$.

\topic{Ad-hoc query}: We mainly focus on developing ad-hoc query evaluation schemes. As of now, the granularity of provenance in our context is at the level of commands executions, the number of activities are constrained by project members' work rate. In case when the \prov\ graph becomes extremely large, indexing techniques and incremental algorithms are more practical. We leave them as future steps.

\subsection{Summarization Operation}
\topic{Alternative Formation} 
We consider alternative of formulation of the summary graph. One way to combine \opseg\ segment graphs is to use context-free graph grammars (CFGG)~\cite{bpql_milo@vldb06} which are able to capture recursive substructures. However without a predefined workflow skeleton CFGG, and due to the workflow noise resulting from the nature of analytics workload, inferring a minimum CFGG from a set of subgraphs is not only an intractable problem, but also possibly leads to complex graph grammars that are more difficult to be understood by the users~\cite{subdue_cfgg@ijait04}. Instead, we view it as a graph summarization task by grouping vertices and edges in the set of segments to a \opsumgraph.

Though requiring all paths in \opsumgraph\ must exist in some segment may look strict and affect the compactness of the result, \opsum\ operator allows using the property aggregation (\propagg) and provenance types (\provtypehop{k}) to tune the compactness of \opsumgraph. Due to the rigidness and the utility of provenance, allowing paths that do not exist in any segment in the summary would cause misinterpretation of the provenance, thus would not be suitable for our context. In situations where extra paths in the summary graph is not an issue, problems with objectives such as minimizing the number of introduced extra paths, and minimizing the description length are interesting ones to be explored further. We leave them as future steps.

\subsection{System Design Decision}
Note the design decision of using a general purpose native graph backend (Neo4j) for high-performance provenance ingestion may not be ideal, as the volume of ingested provenance records would be very large in some applications, e.g., whole-system provenance recording at kernel level~\cite{pass_harvard@atc06,linuxprov_abates@atc15} would generate GBs of data in minutes. The support of flexible and high performance graph ingestion on modern graph databases and efficient query evaluation remain an open question~\cite{dda_microsoft@cidr17}. We leave the issue to support similar operators for general \prov\ graph for our future steps. The proposed techniques in the paper focus on enabling better utilization of the ingested provenance information via novel query facilities and are orthogonal to the storage layer.

\balance

\end{document}